\documentclass[12pt]{article} 
\usepackage[hyperfootnotes=false]{hyperref}
     \usepackage{amsmath}
      \usepackage{amssymb}
  \usepackage{graphicx}
  \usepackage{xcolor}
  \newlength{\abstractwidth}
 \usepackage{mciteplus}

  \newcommand{\be}{\begin{equation}}
  \newcommand{\bea}{\begin{eqnarray}}
  \newcommand{\eea}{\end{eqnarray}}
  \newcommand{\beq}{\begin{equation}}
  \newcommand{\ee}{\end{equation}}
  \newcommand{\eeq}{\end{equation}}
  \newcommand{\N}{{\cal N}}

\def\32{{3 \over 2 } }

\def\diff{\text{diff}}
  \def\ba{\begin{eqnarray}}
  \def\ea{\end{eqnarray}}

 \def\simleq{\; \raise0.3ex\hbox{$<$\kern-0.75em
      \raise-1.1ex\hbox{$\sim$}}\; }
 \def\simgeq{\; \raise0.3ex\hbox{$>$\kern-0.75em
      \raise-1.1ex\hbox{$\sim$}}\; }

\def\R{{\Bbb R}}
\def\Z{{\Bbb Z}}
\def\Ber{\text{Ber}}
\def\M{{\mathcal M}}

\def\sch{\Sch}
\def\Sch{{\rm Sch}}

\begin{document}

\begin{titlepage}
  \bigskip

  \bigskip\bigskip 
 
{\color{white} .}
 \vspace{20pt}

  \bigskip

\begin{center}
{\Large \bf {Fermionic Localization }}\vskip.2cm{\Large\bf{ of the Schwarzian Theory}}
 \bigskip
{\Large \bf { }} 
    \bigskip
\bigskip
\end{center}

  \begin{center}

 \bf {Douglas Stanford  and Edward Witten}
  \bigskip \rm
\bigskip
 
   Institute for Advanced Study,  Princeton, NJ 08540, USA  

  \bigskip \rm
\bigskip
 
\rm

\bigskip
\bigskip

  \end{center}

 \bigskip\bigskip
  \begin{abstract}

The SYK model is a quantum mechanical model that has been proposed to be holographically dual to a $1+1$-dimensional model of a quantum black hole.
An emergent ``gravitational'' mode of this model is governed by an unusual action that that has been called the Schwarzian action.  It governs
a reparametrization of a circle.  We show that the path integral of the Schwarzian theory is one-loop exact.
The argument uses a method of fermionic localization, even though the model itself is purely bosonic.

 \medskip
  \noindent
  \end{abstract}
\bigskip \bigskip \bigskip

  \end{titlepage}

   \tableofcontents

\def\Tr{{\mathrm{Tr}}}
\def\Pf{{\mathrm{Pf}}}

\section{Introduction}
The low-energy description of the SYK model \cite{Sachdev:1992fk,KitaevTalks} and of certain two-dimensional dilaton gravity models \cite{Jackiw:1984je,Teitelboim:1983ux,Almheiri:2014cka} is a theory of a degree of freedom $\phi(\tau)$ that describes a reparametrization of the thermal circle. This theory can be understood as the dynamics of a pseudo-Goldstone mode for the breaking of reparametrization invariance, and it is expected to describe a wide variety of systems with emergent approximate 1d conformal symmetry. The partition function is \cite{KitaevTalks,Maldacena:2016hyu,Jensen:2016pah,Maldacena:2016upp,Engelsoy:2016xyb}
\be\label{schZ}
Z(g) = \int\frac{d\mu[\phi]}{SL(2,\R )}\exp\left[-\frac{1}{2g^2}\int_0^{2\pi} d\tau\left(\frac{\phi''^2}{\phi'^2} - \phi'^2\right)\right].
\ee
The field $\phi(\tau)$ is restricted to be monotone increasing and to wind once around the circle, $\phi(\tau + 2\pi) = \phi(\tau) + 2\pi$, and we mod out by an $SL(2,\R )$ that acts as $f\rightarrow \frac{a f + b}{cf + d}$ where $f = \tan\frac{\phi}{2}$. The $SL(2,\R)$ invariance of the action is made more obvious by writing it as
\be\label{schaction}
I = \frac{1}{2g^2}\int d\tau\left(\frac{\phi''^2}{\phi'^2} - \phi'^2\right) = -\frac{1}{g^2}\int d\tau \left(\Sch(\phi,\tau) + \frac{1}{2}\phi'^2\right) = -\frac{1}{g^2}\int d\tau  \ \Sch\left(\tan\frac{\phi}{2},\tau\right).
\ee
Here
\be\label{schwdef}
\Sch(\phi,\tau)=\frac{\phi'''(\tau)}{\phi'(\tau)}-\frac{3}{2}\left(\frac{\phi''(\tau)}{\phi'(\tau)}\right)^2\ee
is the Schwarzian derivative of a change of coordinates in one dimension.  
We will refer to (\ref{schZ}) as the Schwarzian theory.

In the context of the SYK model, the coupling $g$ is related to the inverse temperature $\beta$, the original SYK coupling $J$, and the number of fermions $N$, by $\frac{1}{g^2}\propto \frac{N}{\beta J}$. At large $N$ and fixed temperature, this is a weakly coupled theory, dominated by small fluctuations about the saddle point $\phi = \tau$. But at very low temperatures $\beta > \frac{N}{J}$ the theory is strongly coupled, and the partition function is dominated by fluctuating configurations of the $\phi(\tau)$ field far from the saddle (see figure \ref{MC} in Appendix \ref{lattice} for pictures). This suggests that the low-energy properties of the SYK model might be difficult to study.

However, the purpose of this paper is to show that the integral (\ref{schZ}) is one-loop exact:\footnote{The work of \cite{Bagrets:2016cdf} already suggested that the Schwarzian theory was exactly solvable. Further strong evidence for one-loop exactness of this integral was reported in \cite{Cotler:2016fpe}. See also the recent work \cite{Bagrets:2017pwq}.}
\be
Z(g) = Z_{\mathrm{one-loop}}(g) = \frac{\#}{g^3}\exp\left(\frac{\pi}{g^2}\right).
\ee 
(Here $\#$ is a constant that depends on regularization but does not depend on $g$.)
As we will see, this fact can be explained by the Duistermaat-Heckman (DH) formula \cite{duistermaat1982variation}, which states that the integral over a symplectic manifold of $\exp(H/g^2)$, where $H$ generates via Poisson brackets a $U(1)$ symmetry of the manifold, is one-loop exact.\footnote{This assertion is actually a special case of a more general localization principle in equivariant cohomology \cite{BV,AB}. (This more general principle has been used, for example, in computing the prepotential of $\N=2$ supersymmetric theories in four dimensions \cite{Nekrasov}.) The statement that the path integral is one-loop exact assumes that the fixed
points of the $U(1)$ symmetry are isolated.  The general statement involves an integral over this fixed point set.
There is also a generalization that involves localization by nonabelian symmetries \cite{WittenOldpaper,JK,CMR}. We will not need any of
these generalizations here.} To apply the DH formula to the integral (\ref{schZ}), we need two facts. First, the space over which we integrate in (\ref{schZ}) is $\text{diff}(S^1)/SL(2,\R )$, which is symplectic.
 And second, the Schwarzian action generates a $U(1)$ symmetry of this manifold that corresponds to time translations. Together, these two facts explain the one-loop exactness.

The Duistermaat-Heckman formula may be unfamiliar to some readers, so we will sketch a physics proof (along the lines of \cite{Bismut:1986oha,WittenOldpaper}). Suppose given a symplectic manifold $M$ with coordinates $x^i$, and with a Hamiltonian $H$ that generates a $U(1)$ symmetry via
\be
\delta x^i  = v^i = \omega^{ij}\partial_j H.
\ee
Here, $\omega^{ij}$ is the inverse of the symplectic form $\omega_{ij}$. We would like to evaluate the integral
\be\label{dhZ}
Z = \int d^nx d^n\psi \exp\left[\frac{1}{2}\omega_{ij}\psi^i\psi^j + \frac{H(x)}{g^2}\right].
\ee
Here the $\psi^i$ are Grassmann variables that transform  under diffeomorphisms of $M$ as the differentials $d x^i$, so that there is a natural measure $d^nx d^n\psi$.
Integrating out the $\psi$'s in eqn. (\ref{dhZ})  gives the symplectic measure $\mathrm{Pf}(\omega)d^nx$ for the integral over the $x$ coordinates (here $\mathrm{Pf}(\omega)$ is the Pfaffian\footnote{The Pfaffian is the square root of the determinant, but for an antisymmetric matrix such as $\omega$, it is more natural to speak of the Pfaffian.}
of the antisymmetric matrix $\omega_{ij}$).  This is the usual symplectic measure on the symplectic manifold $M$. The point of writing the measure this way is that (\ref{dhZ}) has a fermionic symmetry
\be
Qx^i = \psi^i, \hspace{20pt} Q \psi^i = v^i
\ee
that allows us to analyze the integral by  localization. To check the fermionic symmetry, one uses the fact that the symplectic form is closed, $d\omega = 0$. One can further check that $Q^2$ acts as the generator  of the $U(1)$ symmetry. Accordingly, $Q^2=0$ in acting on $U(1)$-invariant operators or functions.
It follows that we can add $QV$ to the action, for any $U(1)$-invariant $V$, without changing the integral. To prove the DH formula, we choose $V = g_{ij}v^i\psi^j$ where $g_{ij}$ is any $U(1)$-invariant metric. Then we add $QV$ to the action with a large coefficient. This localizes the integral to the critical points of $H$, along with the correct measure for the one-loop integral about those points. This establishes the one-loop exactness. We give some more details on this proof in a slightly more general setting in Appendix \ref{appendixDH}.

The plan of this paper is as follows. In section two, we will discuss the case of the Schwarzian theory in more detail, including the the symplectic form, the associated measure, perturbation theory, and the exact partition function. In section three, we will give further details on the relationship of the Schwarzian theory to the coadjoint orbit of Virasoro, and discuss some generalizations. In particular, we discuss the exact partition function and density of states of the $\mathcal{N} =1 $ and $\mathcal{N} = 2$ super-Schwarzian theories \cite{Fu:2016vas}. In Appendix A, we describe the extension of the DH formula
to a supermanifold.  In Appendix B, we describe a lattice regularization of the Schwarzian theory and some sample monte carlo configurations.
Finally, in Appendix C, we discuss some simple correlation functions in the Schwarzian theory.

\section{The Schwarzian Theory}

The localization argument for the Schwarzian theory relies on two facts:  the space we are integrating over is symplectic, and the action generates a $U(1)$ symmetry of that space. We will explain both of these facts in this section, before showing mechanically that the two-loop term vanishes and discussing the exact one-loop answer.

\subsection{The Symplectic Form and the Hamiltonians}\label{symham}
The space that we integrate over in the path integral (\ref{schZ}) is the symplectic manifold $\text{diff}(S^1)/SL(2,\R )$. As a first step, we should understand the symplectic form for this space and the integration measure that follows from it. The form $\omega$ can be derived from the fact that $\text{diff}(S^1)/SL(2,\R )$ is the orbit of a particular coadjoint vector under the action of the Virasoro group \cite{kirillov1981orbits,Witten:1987ty,Bakas:1988bq}, and coadjoint orbits are always symplectic. See section \ref{sthree} for more details.
We also give a direct explanation momentarily.

We can label points on $\text{diff}(S^1)/SL(2,\R)$ by elements $\tau(\phi)$ of $\text{diff}(S^1)$, with the understanding that we are identifying by a right action of $SL(2,\R)$ as $f\equiv \tan\frac{\phi}{2}\rightarrow \frac{af+b}{cf+d}$. In the application to the Schwarzian theory, it turns out to be somewhat more convenient to think in terms of the inverse $\text{diff}(S^1)$ element, which we write as $\phi(\tau)$. This is the same variable that appears in the action (\ref{schaction}). With this choice of coordinate, the symplectic form is given by \cite{Witten:1987ty,Alekseev:1988ce}
\begin{align}\label{omega}
\omega &= \int_0^{2\pi}d\tau\left[\left(\frac{d\phi(\tau)}{\phi'(\tau)}\right)'\wedge\left(\frac{d\phi(\tau)}{\phi'(\tau)}\right)'' - 2\,\Sch\left(\tan\frac{\phi}{2},\tau\right)\left(\frac{d\phi(\tau)}{\phi'(\tau)}\right)\wedge \left(\frac{d\phi(\tau)}{\phi'(\tau)}\right)'\right]\\
&=\int_0^{2\pi} d\tau \left[\frac{d\phi'(\tau)\wedge d\phi''(\tau)}{\phi'^2(\tau)} - d\phi(\tau)\wedge d\phi'(\tau)\right].\label{omega2}
\end{align}

This formula probably requires some explanation.  First of all, here $d$ is an abstract exterior derivative that acts on the field $\phi$ but not on
the spatial coordinate $\tau$ (or on a function of $\tau$).  One can think of $\phi(\tau)$ as a bosonic function in one dimension and $\psi(\tau)=d\phi(\tau)$
as a fermion.  As $d$ does not act on $\tau$, likewise it commutes with $\partial_\tau$, so for instance $d\phi'=d(\partial_\tau\phi)$ is the same as
$\partial_\tau d\phi$.  Using such facts, we  can rewrite (\ref{omega2}) in the form
\be\label{omega3} \omega=\int_0^{2\pi}d\tau\left( \left(\frac{d\phi'}{\phi'}\right)\partial_\tau\left(\frac{d\phi'}{\phi'}\right) -d\phi\partial_\tau d\phi\right).\ee
We have omitted the explicit wedge product symbol -- which one can think of as a reminder that $d\phi(\tau)$ is a fermionic field -- and we have
added a term that vanishes because $(d\phi')^2=0$ by fermi statistics.

Now let us describe and verify the important properties of $\omega$.   The first basic property is that it is closed, $d\omega=0$.  This is actually manifest in eqn.
(\ref{omega3}), which has been written entirely in terms of $d\phi$ and $d\phi'/\phi'=d\log\phi'$  (and their $\tau$ derivatives), both of which are exact and
in particular closed.

The second basic property of $\omega$ is that it is invariant under $\diff\,S^1$ acting on $\tau$.  An infinitesimal $\diff\,S^1$ transformation $\delta \tau=\alpha(\tau)$
acts on $\phi$ by 
\be\label{umn}\delta \phi(\tau)=\alpha(\tau)\phi'(\tau). \ee
This corresponds to a vector vield $V_\alpha$ on the space of $\phi$'s.  We want to know if $V_\alpha$ leaves $\omega$ invariant.

In general, if $\Omega$ is any differential form and $V$ is any vector field, the condition that $V$ leaves $\Omega$ fixed is that $(d\iota_V+\iota_V d)\Omega=0$.
Here $\iota_V$ is the operation of contracting $V$ with  $\Omega$.  If $d\Omega=0$, the condition reduces to $d(\iota_V\Omega)=0$.
We are in that situation because $d\omega=0$, so to show that $\omega$ is $V_\alpha$-invariant, we have to show that $d(\iota_{V_\alpha}\omega)=0$.
We will do this by finding a function $H_\alpha$ such that 
\be\label{defham}\iota_{V_\alpha}\omega=d H_\alpha.\ee

These functions have a special interpretation (or more exactly will have such an interpretation after we remove zero-modes and make $\omega$
nondegenerate). If $\omega$ is a symplectic form, then eqn. (\ref{defham}) means\footnote{In terms of indices, this equation reads $V_\alpha^i\omega_{ij}=\partial_j H_\alpha$.
This is equivalent to $V_\alpha^i=(\omega^{-1})^{ij}\partial_j H_\alpha$, which may be a more familiar version of the formula.} that  $H_\alpha$ is the Hamiltonian function that via Poisson brackets generates the infinitesimal transformation
(\ref{umn}) of $\phi$.  In particular, on setting $\alpha=1$, we will get the  Hamiltonian function $H$ associated to the ``time'' translation symmetry
 $\delta\phi=\phi'$. It will turn out that this function (up to a factor of $-1/g^2$) is the Schwarzian action (\ref{schaction}).  That fact is the basic reason
 that the Duistermaat-Heckman formula is relevant to the Schwarzian theory.  
   
 Concretely, $\iota_{V_\alpha}\omega$ is computed by replacing  one copy of $d\phi$ in eqn. (\ref{omega3}) with $\delta\phi=\alpha \phi'$.   One makes
 this replacement for any one copy of $d\phi$, but one has to take account of signs: $\iota_{V_\alpha}$ is regarded as a fermionic operator that anticommutes   
 with $d\phi$, so we include a plus or minus sign if it acts on the leftmost or rightmost factor of $d\phi$.   To make this explicit, let us write $\omega=\omega_1
 +\omega_2$, with $\omega_2=-\int_0^{2\pi}d\tau \,d\phi\partial_\tau d\phi$ and $\omega_1=\omega-\omega_2$.   
 Then $\iota_{V_\alpha}\omega_2=-2\int_0^{2\pi}d\tau \,\alpha\phi' \partial_\tau d\phi
 =-2\int_0^{2\pi}d\tau \,\alpha \phi' d\phi'$.  But this is $d H_{\alpha,2}$ with $H_{\alpha,2}=-\int_0^{2\pi}d\tau \,\alpha \,\phi'^2$.  
 
Similarly, 
\be\label{zota}\iota_{V_\alpha}\omega_1=2\int_0^{2\pi}d\tau\,\left(\frac{\partial_\tau(\alpha\phi')}{\phi'} \partial_\tau\frac{d\phi'}{\phi'}\right)=2\int_0^{2\pi}
d\tau\left(\left(\alpha'   +\alpha \frac{\phi''}{\phi'} \right)\partial_\tau \frac{ d\phi'}{\phi'}\right). \ee 
But this is $dH_{\alpha,1}$ with $H_{\alpha,1}=\int_0^{2\pi}d\tau\,\left(2\alpha' \frac{\phi''}{\phi'}+\alpha \left(\frac{\phi''}{\phi'} \right)^2\right)$.
Here we use $\partial_\tau (d\phi'/\phi')=d(\phi''/\phi')$.

Finally, then, the Hamiltonian function that generates the $\diff\,S^1$ transformation $\delta\phi=\alpha\phi'$ will be
\be\label{mota}H_\alpha=\int_0^{2\pi}d\tau \left( \alpha\left(\left(\frac{\phi''}{\phi'}\right)^2-(\phi')^2\right)+2\alpha' \frac{\phi''}{\phi'}\right) = -2\int_0^{2\pi}d\tau\, \alpha\,\sch\left(\tan\frac{\phi}{2},\tau\right).\ee
Setting $\alpha=1$, we learn that the ordinary Hamiltonian that generates $\delta\phi=\phi'$ will be the Schwarzian action (\ref{schaction}), up to a factor of $1/2g^2$.

So far, we have shown that $\omega$ is closed and $\diff\,S^1$ invariant and we have identified what will be the Hamiltonian functions.  
We still have to explain in what sense $\omega$ can be interpreted as a symplectic form.
Since $\phi(\tau)$ is monotone increasing and winds once around the circle, it can be interpreted as an element of $\diff\,S^1$, mapping $\tau$ to $\phi(\tau)$.  The form $\omega$ is a closed
form on $\diff\,S^1$, but is not symplectic because it is degenerate, that is it has zero-modes.  To see the zero-modes in the most direct way, let us expand around $\phi=\tau$,
which we think of as the identity element of $\diff\,S^1$, with $\phi=\tau+\sum_n u_n e^{in \tau}$.  At $u_n=0$,  one finds that
$\omega=(-2\pi i) \sum_{n\in \Z} (n^3-n)d u_n d u_{-n}$.  The function $n^3-n$ vanishes precisely for $n=-1,0,1$, so at the identity element of $\diff\,S^1$, $\omega$
has precisely three zero-modes, that is, it has a three-dimensional kernel.  Since $\omega$ is $\diff\,S^1$-invariant, the same is true everywhere.

The three zero-modes have a simple interpretation.   For $n=1,0,-1$, let $W_n$ be the vector field
\be\label{vecf}\delta\phi =e^{in\phi}. \ee
They generate an action of $SL(2,\R)$ on $\phi$, and this action commutes with the $\diff\,S^1$ action generated by the $V_\alpha$.   One may ask if $\omega$
is invariant under this second action of $SL(2,\R)$.  The criterion is again that $d\iota_{W_n}\omega=0$.  But here a short calculation with some integration by parts
reveals that more is true:
$\iota_{W_n}\omega=0$.  Thus $W_n$ are the three zero-modes of $\omega$.  $SL(2,\R)$ invariance of $\omega$ would require the weaker condition $d(\iota_{W_n}\omega)=0$.
The fact that actually $\iota_{W_n}\omega=0$ even without acting with $d$ means that $\omega$ is not just $SL(2,\R)$-invariant but is a ``pullback'' from a quotient space
in which one divides by $SL(2,\R)$.

Let us recall that if $G$ is a nonabelian group, then there is a left and right action of $G$ on itself.  The left action on $g\in G$ is by $g\to fg$, and the right action is by
$g\to g h^{-1}$, where here $f,h\in G$.  It is a matter of convention which is called the ``left'' action and which the ``right'' action, as the two are exchanged by $g\to g^{-1}$.
In the case at hand, the group of interest is $\diff\,S^1$, parametrized by $\phi$.  The ``left'' action is generated by $\delta\phi=\epsilon(\tau)\partial_\tau\phi$.
It commutes with the ``right'' action, which is generated by $\delta\phi=g(\phi)$ for an arbitrary function $g(\phi)=\sum_{n\in \Z}g_n e^{in\phi}$.    What we have found
in the last paragraph is that $\omega$ is a pullback from $\diff\,S^1/SL(2,\R)$, where $SL(2,\R)$ acts on the right as a subgroup of $\diff\,S^1$.  Differently put, we can view $\omega$ as a closed two-form on the
quotient space $\diff\,S^1/SL(2,\R)$.  If we do this, since we have removed the zero-modes, $\omega$ becomes nondegenerate and therefore is a symplectic form.
The quotient space $\diff\,S^1/SL(2,\R)$ admits only one action of $\diff\,S^1$, which descends from the left action of $\diff\,S^1$ on itself.  When viewed as a form 
on the quotient space, $\omega$ is $\diff\,S^1$-invariant, since it is left-invariant as a form on the group manifold.

In general, given a symplectic manifold, once one picks  coordinates $u_1,\dots,u_n$,
the symplectic form becomes an antisymmetric matrix and the  measure in that coordinate
system is $d^nu$ times the Pfaffian of the matrix. In the present context, we use the function $\phi(\tau)$ (subject to a gauge condition to remove
zero-modes) to parametrize $\diff\,S^1/SL(2,\R)$.  In this parametrization, the symplectic measure is not just the ordinary path integral
measure $D\phi$, but is this times the Pfaffian of the symplectic form.  From a physical point of view, the evaluation of the Pfaffian is
more understandable 
 if we rename $d\phi(\tau)$ as a fermion field $\psi(\tau)$.  Then $\omega$, being bilinear in $\psi$,
is a quadratic fermion action, and the Pfaffian of $\omega$ is just the path integral for $\psi$ with this action.   We turn next
to an analysis of that path integral.

\subsection{Evaluation of the Measure}
Naively, it seems that the Pfaffian of (\ref{omega}) will give a nonlocal measure for $\phi$. However, this depends on the choice of gauge that is used to remove
$SL(2,\R)$ zero-modes.  If we pick a local gauge condition, like $\phi(0) = 0$, $\phi'(0) = 1$, $\phi''(0) = 0$, then the measure actually turns out to be local, and equal to the measure on $\text{diff}(S^1)/SL(2,\R )$ inherited from the measure on $\text{diff}(S^1)$ discussed in \cite{Bagrets:2016cdf}:
\be\label{naiveM}
d\mu_{\text{diff}(S^1)}[\phi] = \prod_\tau \frac{d\phi(\tau)}{\phi'(\tau)}.
\ee

Let us show in some detail how to derive this from the symplectic form. It is convenient to work with the expression for $\omega$ on the first line of (\ref{omega}). To compute the Pfaffian we view this as (minus two times) an action for a periodic fermion variable $\psi = d\phi(\tau)$. We can simplify the action somewhat by defining a new fermion variable $\eta = \psi/\phi'$. The change of variables from $\psi$ to $\eta$ gives us the infinite product of $1/\phi'(\tau)$ in (\ref{naiveM}) (minus three factors  for the gauge-fixing). This looks promising, but we have a leftover 
fermionic integral that seems to depend on $\phi(\tau)$:
\be\label{pfoft}
\int D\eta\, \eta(0)\eta'(0)\eta''(0)\exp\left[\frac{1}{2}\int d\tau \Big(\eta'\eta'' - T(\tau)\eta\eta'\Big)\right], \hspace{20pt} T(\tau) \equiv 2\,\Sch\left(\tan\frac{\phi}{2},\tau\right).
\ee
Remarkably, (\ref{pfoft}) is actually independent of $\phi(\tau)$. To show this, we evaluate the path integral by representing it as a trace in a Hilbert space. In other words, we canonically quantize the $\eta$ action. We can construct an equivalent action  that is linear in $\tau$ derivatives
by ``integrating in'' some additional fields:
\begin{align}\label{fot}
\int D\eta\, &\eta(0)\eta'(0)\eta''(0)\exp\left[\frac{1}{2}\int d\tau \Big(\eta'\eta'' - T(\tau)\eta\eta'\Big)\right]\\
&=\int {D}\eta\,{D}w\,{D}b \,{D}c\, c(0)\eta(0)b(0)b'(0)\exp\left[\frac{1}{2}\int d\tau\Big(w(\eta'-b) + b b' - T(\tau)\eta b -cc'\Big)\right]. \notag
\end{align}  (All fermion fields here are assumed to be periodic on the circle.)
Integrating out $w$ sets $b = \eta'$.  The action in (\ref{fot}) then reduces to the previous action in (\ref{pfoft}) and the fermion insertions $\eta(0)b(0)b'(0)$ reduce to the insertions $\eta(0)\eta'(0)\eta''(0)$
in (\ref{pfoft}). In order to have an even number of Majorana fermions, we have included in (\ref{fot}) a fourth decoupled field $c$ (and an insertion of $c(0)$ to make the periodic path integral nonzero).  For future reference, note that the equation of motion for $b$ gives $2b'+w+T(\tau)\eta=0$, so $c(0)\eta(0)b(0)b'(0)$
will be equivalent as an operator acting on physical states to $-\frac{1}{2} c(0)\eta(0)b(0)w(0)$:
\be\label{useeq}c(0)\eta(0)b(0)b'(0)\sim -\frac{1}{2} c(0)\eta(0)b(0)\left[w(0) +T(0)\eta(0)\right] = -\frac{1}{2} c(0)\eta(0)b(0)w(0).\ee
In the final step we used that $\eta(0)^2 = 0$.

To understand the Hilbert space for $w,b,\eta,c$, it is convenient to write the fields in terms of four canonical Majorana fermions
\be
\eta = \chi_1 + i\chi_2, \hspace{20pt} w = -\chi_1 + i\chi_2, \hspace{20pt} b = i\chi_3, \hspace{20pt} c = \chi_4.
\ee
Then the exponential  in (\ref{fot}) becomes
\be\label{readoff}
\exp\left[-\frac{1}{2}\int d\tau\Big(\sum_j\chi_j\chi_j' + i\chi_3(\chi_1-i\chi_2) - iT(\tau)\chi_3(\chi_1+i\chi_2)\Big)\right].
\ee
This theory is now easy to quantize in a four-dimensional Hilbert space, setting
\be
\chi_1 = \frac{1}{\sqrt{2}}X\otimes I, \hspace{20pt} \chi_2 = \frac{1}{\sqrt{2}}Y\otimes I, \hspace{20pt} \chi_3 = \frac{1}{\sqrt{2}}Z\otimes X, \hspace{20pt} \chi_4 = \frac{1}{\sqrt{2}}Z\otimes Y,
\ee
where $X,Y,Z$ are the standard Pauli matrices and $I$ is the identity. Then our path integral is a representation of the time evolution operator for a Euclidean quantum mechanics problem with a time-dependent (and non-Hermitian) Hamiltonian that we read off from (\ref{readoff}):
\be
H(\tau)= \frac{1}{2}\left[ i\chi_3(\chi_1-i\chi_2) - iT(\tau)\chi_3(\chi_1+i\chi_2)\right] = -\frac{i}{2}\left( \begin{array}{cccc}
0 & 0 & 0 & T(\tau)  \\
0 & 0 & T(\tau) & 0\\
0 & 1 & 0 & 0\\
1 & 0 & 0 & 0 \end{array} \right).
\ee
We can now compute the operator $U(2\pi) = P e^{-\int_0^{2\pi}H(\tau)d\tau}$ that describes evolution all the way around the circle, by solving the Schrodinger equation $\Psi' = -H \Psi$.
That equation is equivalent to a pair of equations
\be
\frac{d}{d \tau}\begin{pmatrix} \psi_1\cr \psi_2\end{pmatrix} = \frac{i}{2}\begin{pmatrix} 0 & T(\tau)\cr 1&0\end{pmatrix}\begin{pmatrix} \psi_1\cr \psi_2\end{pmatrix} \ee
for two-component wavefunctions.
Eliminating the bottom components, this equation
is equivalent to the second order equation
\be
\partial_\tau^2\Psi + \frac{1}{2}\Sch\left(\tan\frac{\phi}{2},\tau\right)\Psi = 0.
\ee
This equation has two linearly independent solutions $\cos(\phi/2)/\sqrt{\phi'}$ and $\sin(\phi/2)/\sqrt{\phi'}$, both of which satisfy $\Psi(\tau+2\pi) = -\Psi(\tau)$. We conclude that the operator
that describes time evolution around the circle is simply $U(2\pi) = -1$. The fact that this is independent of $\phi$ is the crucial point that underlies the fact that the path integral (\ref{pfoft}) is independent of $\phi$. 

Explicitly, the desired path integral (\ref{pfoft}) is $\Tr\,[(-1)^F U(2\pi) (-1/2)c\eta bw(0)] = 1$, where the factor $(-1)^F=Z\otimes Z$ expresses the fact that the fermions obey periodic boundary conditions, and
the insertions $(-1/2)c\eta bw(0) = -\frac{1}{4}(1 +Z)\otimes Z$ are the original fermion zero-mode insertions of eqn. (\ref{fot}),
expressed in eqn. (\ref{useeq}) in a Hamiltonian language. It follows that the measure reduces  to the naive measure (\ref{naiveM}).\footnote{There were various arbitrary choices of sign in this calculation, involving the ordering of fermion zero-mode insertions and the sign of the measure for the fermion zero-modes.  In general, on a symplectic manifold $M$ with symplectic form $\omega$, the sign of $\Pf(\omega)$ depends on the
orientation of $M$.  Conventionally, $M$ is oriented to make $\Pf(\omega)$ positive, and thus we have chosen
signs to get a result $+1$ rather than $-1$ in the above calculation.  In any case, the path integral measure that
we want is positive.}

\paragraph{Why This Measure?}

In general, if $G$ is any group, it acts on itself on both the left and the right, as we recalled in section \ref{symham}.  There is always
(up to a constant multiple) a unique left-invariant measure $\Omega_\ell$ on $G$, obtained by picking an arbitrary measure on the tangent
space to the identity in $G$ and then translating it over the group manifold so as to be left-invariant.  Under a mild condition on $G$, which is
satisfied for $\diff\,S^1$, a left-invariant measure is also right-invariant and it does not matter if we ask for a measure to be left- or right-invariant.\footnote{Let $h$ be any element of $G$, acting on $G$ on the right.  As the right action of $G$ on itself commutes with the left action, $h$
will transform a left-invariant measure $\Omega_\ell$ to another left-invariant measure $\Omega'_\ell$.  But as the left-invariant measure
is unique up to a real multiple, $h$ actually transforms $\Omega_\ell$ to $u(h)\Omega_\ell$ for some real constant $u(h)$.  The map $h\to u(h)$
is a one-dimensional real representation of $h$.  Thus if $G$ has no nontrivial real one-dimensional representations -- as is the case for
$\diff\,S^1$ -- $u(h)$ will be identically 1 and a left-invariant measure is automatically right-invariant. (In this case, the group $G$ is said
to be unimodular.)}   In the case of $\diff\,S^1$,
the convention adopted in section \ref{symham} was that the left action is  $\phi(\tau)\rightarrow \phi(g(\tau))$ and the right action is  $\phi(\tau) \rightarrow h(\phi(\tau))$.

The symplectic description of the measure is most natural on the quotient space $\diff\,S^1/SL(2,\R)$.  Here, since we divide by $SL(2,\R)$ acting
on the right, the only natural group action that remains is the one that descends from the left action of $SL(2,\R)$ on itself.  However,
the same measure can be constructed by starting with an invariant measure on $\diff\,S^1$ -- as just explained, it does not matter if
one takes this to be a left-invariant or right-invariant measure -- and then dividing by $SL(2,\R)$ (itself endowed with an invariant
measure).   This point of view about the measure was adopted in \cite{Bagrets:2016cdf}.  The starting point there was the measure (\ref{naiveM})
on the group manifold.  From that point of view, the right invariance of the measure on $\diff\,S^1$ seems more essential, because it is this
that enables one to divide by $SL(2,\R)$ and get a measure on the quotient.

Ideally, one would like to derive the symplectic or quotient measure directly from a theory where the Schwarzian action arises, such as SYK or dilaton gravity. We have not done this, but we will make two further comments that suggest that this choice is reasonable. 

First, the formula (\ref{naiveM}) makes it clear that the measure is UV-divergent, since it involves a product of $\phi'$ at each point. If we were to write this as a term in the action, we would have $S_{measure} = \delta(0)\int d\tau \log \phi'$. As we will see, this UV divergence is actually necessary to cancel divergences that appear in the naive perturbation theory of the Schwarzian action.\footnote{See \cite{Brezin:1976ap} for a similar phenomenon in the $O(N)$ sigma model. As in that setting, one can change variables in the Schwarzian theory to variables that simplify the measure and make the absence of UV divergences manifest, by choosing $\Phi = (\tan \frac{\phi}{2})'$. The measure (\ref{naiveM}) is simply flat in $\Phi$. This was pointed out in \cite{Bagrets:2016cdf}, where it is shown that the Schwarzian action is also simple in this coordinate. Unfortunately, it seems to become more complicated when we impose $SL(2,\R )$ gauge-fixing conditions, and we have not found a good way to use this $\Phi$ variable.}

Second, we note here that the one-loop exact answer for the partition function that we will get from (\ref{Zwithpsi}) agrees precisely with the answer in the triple-scaled limit of SYK \cite{Cotler:2016fpe} that is conjectured to isolate the pure Schwarzian theory. This suggests that at least in that limit, SYK produces the symplectic measure.

\subsection{Perturbation Theory}
In this section we show how to do perturbation theory for the Schwarzian action including the symplectic measure. We will evaluate the one-loop answer (which is the exact answer by the DH formula) and we will show explicitly that the two-loop term vanishes.

The integral of the Schwarzian action with the measure $\Pf(\omega)$ can be written as
\be\label{Zwithpsi}
Z(g) = \int \frac{\mathcal{D}\phi\mathcal{D}\psi}{SL(2,\R)}\exp\left\{-\frac{1}{2}\int_0^{2\pi} d\tau\left[\frac{\phi''^2}{g^2\phi'^2}  - \frac{\phi'^2}{g^2} + \frac{\psi''\psi'}{\phi'^2} - \psi' \psi\right]\right\}.
\ee
Here, we have introduced a periodic Grassmann field $\psi$ to write the Pfaffian of the symplectic form (\ref{omega2}) as a path integral. The formula (\ref{Zwithpsi}) is analogous to the expression (\ref{dhZ}) from the finite-dimensional setting.

We need to be careful about the space of fields over which we are integrating. As described above, the subtlety is due to the fact that the $\phi$ coordinate is a coordinate on $\text{diff}(S^1)$, whereas we want to integrate only over $\text{diff}(S^1)/SL(2,\R )$. For perturbation theory, a convenient choice of coordinates on the quotient space is to gauge-fix
\begin{align}\label{gaugeCond}
\int d\tau \,\varepsilon(\tau) = \int d\tau \,e^{\pm i\tau}\varepsilon(\tau) = 0, \hspace{20pt} 
\int d\tau\,\psi(\tau) = \int d\tau\, e^{\pm i\tau}\psi(\tau) = 0,
\end{align}
where we define $\varepsilon$ by $\phi(\tau) = \tau + g\varepsilon(\tau)$. Notice that $\varepsilon$ is a strictly periodic variable.

We will consider the first few orders of perturbation theory in $g$: the classical answer, the one-loop determinant, and the two-loop term. With our gauge conditions, the unique classical solution that satisfies $\phi(\tau+2\pi) = \phi(\tau)$  is simply $\phi(\tau) = \tau$ or equivalently $\varepsilon = 0$. The action for this configuration is $-\frac{\pi}{g^2}$. The expansion of the action about this solution in powers of $g$ is
\begin{align}
-I = \frac{\pi}{g^2} - \frac{1}{2}\int_0^{2\pi} du\Big[ &\varepsilon''^2  - \varepsilon'^2 + \psi'' \psi' - \psi' \psi + g\left(-2\varepsilon'\varepsilon''^2 - 2\varepsilon'\psi'' \psi'\right)\notag\\&+g^2\left(3\varepsilon'^2\varepsilon''^2 + 3 \varepsilon'^2\psi'' \psi'\right)+ O(g^3)\Big].
\end{align}
The quadratic action for $\varepsilon$ and $\psi$ is independent of $g$, so it looks like the one-loop term should just give a constant. However, we have a factor of $g$ for each fourier mode of $\varepsilon$ from the transformation of the measure $d\phi = g\, d\varepsilon$. The product over all fourier modes would give something $g$-independent, after regularization. But we are fixing three of the fourier modes in (\ref{gaugeCond}), so the correct regularized answer for the determinant is proportional to $g^{-3}$. We conclude that 
\be
Z_{one-loop}(g) = \frac{\#}{g^3}\exp\left(\frac{\pi}{g^2}\right).
\ee
This is the same conclusion that was reached in \cite{Maldacena:2016hyu}.

To do the two-loop computation we use the propagators\footnote{The exact form of $G(\tau)$ will not be needed, but it is given by 
\be
G(\tau) = \frac{1}{2\pi}\sum_{|n|\ge 2}\frac{e^{-in\tau}}{n^2(n^2-1)} = \frac{1}{2\pi}\left[-\frac{(\tau-\pi)^2}{2} + (\tau-\pi)\sin(\tau) + \frac{5}{2}\cos(\tau) + 1 + \frac{\pi^2}{6}\right]
\ee
for $0\le \tau\le 2\pi$, and extended by periodicity outside this range. Note that $G$ is an even function.}
\begin{align}
\langle \varepsilon(\tau)\varepsilon(0)\rangle &= G(\tau), \hspace{20pt} \langle \psi(\tau)\psi(0)\rangle = G'(\tau).
\end{align}
The two-loop term in $Z$ is the order $g^2$ contribution. This comes from either expanding down the $g^2$ term in the action once, or expanding down the $g$ term twice. The former gives
\be
-\frac{3g^2}{2}\int d\tau\left[\langle \varepsilon'^2\varepsilon''^2\rangle  + \langle \varepsilon'^2\psi''\psi'\rangle\right] = -\frac{3g^2}{2}\left[2G'''(0)^2 - G''(0)G''''(0) + G''(0)G''''(0)\right].
\ee
The first term is zero because $G$ is an even function.\footnote{This is slightly subtle, because $G$ has a singularity at $\tau = 0$ and $\lim_{\tau\rightarrow 0^+} G'''(\tau)$ is not zero. Our manipulations make sense if we replace $G$ with a function that smooths out the singularity in a symmetric way. Equivalently, we can put a cutoff on the absolute values of the frequencies that we consider.} The second and third terms (which come respectively from $\langle \varepsilon'^2\rangle\langle \varepsilon''^2\rangle$ and $\langle\varepsilon'^2\rangle\langle \psi''\psi'\rangle$) cancel. The second two-loop term, which we get by expanding the $O(g)$ term down twice, gives
\begin{align}
\frac{g^2}{2}\int d\tau_1d\tau_2&\left[\langle \varepsilon'\epsilon''(\tau_1)^2 \varepsilon'\varepsilon''(\tau_2)^2\rangle + \langle \varepsilon'\psi''\psi'(\tau_1)\varepsilon'\psi''\psi'(\tau_2)\rangle\right] \\
&= \frac{g^2}{2}\int d\tau_1d\tau_2\Big[-2G''G''''^2 -4 G'''^2 G'''' + G''G''''^2 - G''G'''G'''''\Big],
\end{align}
where all $G$ are $G(\tau_1-\tau_2)$. We can integrate the last term by parts to $G'''^2 G'''' + G'' G''''^2$. The expression inside the integral then reduces to $-3G'''^2G''''$, which is a total derivative. So we conclude that the entire two-loop contribution vanishes.

\subsection{The Exact Partition Function and Density of States}\label{exact}
The proof of the DH formula that we sketched in the Introduction uses a supersymmetric localization argument that is proved by constructing an extended
action with fermion fields and a suitable fermionic symmetry.

We would like to implement this procedure for the reparametrization field $\phi(\tau)$ that appears in the Schwarzian action.
Here we have to take care of the following detail.  The procedure described in the introduction makes sense for any symplectic manifold $M$ with a $U(1)$
 symmetry.  However, the field $\phi(\tau)$ without any further condition or equivalence
 would represent an element of $\diff\,S^1$, which is not a symplectic manifold.  
 The localization procedure that we are looking for is only possible because in the Schwarzian theory, $\phi(\tau)$ is properly viewed as an element of
 the quotient space $\diff\,S^1/SL(2,\R)$, which is symplectic.
 
A related fact is the following.  To show the one-loop exactness of the path integral for the Schwarzian action, we will want the classical solution $\phi(\tau)=\tau$
to be $U(1)$-invariant, where $U(1)$ acts by shifting $\tau$ by a constant.  This is clearly not true if we regard $\phi(\tau)$ as an element of $\diff\,S^1$, since
$\phi(\tau)=\tau$ is not invariant under adding a constant to the right hand side.
But if $\phi$ is regarded as an element of $\diff\,S^1/SL(2,\R)$ (so that in particular $\phi$ is considered equivalent to $\phi+c$ for constant $c$)
then $\phi(\tau)=\tau$ is indeed $U(1)$-invariant.\footnote{For both of the issues we have raised, it would be just as good if $\phi$ were regarded
as an element of $\diff\,S^1/U(1)$, where $U(1)$ acts by a constant shift of $\phi$.  Here $\diff\,S^1/U(1)$ is a symplectic
manifold, and if $\phi$ is regarded as an element of $\diff\,S^1/U(1)$, then $\phi(\tau)=\tau$ is invariant under a constant shift of $\tau$.}

Therefore in order to implement the localization procedure, we have to view $\phi$ as an element of $\diff\,S^1/SL(2,\R)$.  A convenient way to do this is
to write 
$\phi=\tau+g\varepsilon$, where $\varepsilon$ is required to satisfy the constraints (\ref{gaugeCond}), and add a fermion field $\psi$ that satisfies the same
constraints.  One can then define the fermionic symmetry 
\be Q\varepsilon = \psi, ~~~~Q\psi=\varepsilon', 
\ee
whose square is the $U(1)$ generator corresponding to $d/d\tau$, in keeping with the general framework.  
To localize the path integral, we add $QV$ to the action with a large coefficient, where for example
\be
V = \int d\tau \psi \varepsilon', \hspace{20pt} QV = \int d\tau\left[\varepsilon'^2 + \psi'\psi\right].
\ee
Adding this term with a large coefficient $s$ does not  affect the one-loop or classical terms in the partition function, but because it makes the separated-point propagators small it suppresses higher-loop corrections. By the logic of the DH proof, the partition function is independent of $s$, so all loop
contributions must actually vanish. We conclude that
\be
Z(g) = Z_{one-loop}(g) = \frac{\#}{g^3}\exp\left(\frac{\pi}{g^2}\right).
\ee

\def\H{{\mathcal H}}
\paragraph{A Note On The Density Of States}
It is interesting to consider the density of states that gives rise to this partition function. In the application to SYK or dilaton gravity, we have that $g^2 = \beta/(2\pi C)$, where $\beta$ is the inverse temperature, and $C$ is a coefficient with dimensions of length. So in terms of $\beta$, the partition function is
\be\label{partbeta}
Z(\beta) \propto \frac{1}{\beta^{3/2}}\exp\left(\frac{2\pi^2 C}{\beta}\right).
\ee
This is equal to $\int_0^{\infty}\rho(E)e^{-\beta E}d E$   with
\be
\rho(E) \propto \sinh\left(2\pi\sqrt{2 CE}\right).
\ee

It is tempting to call $\rho(E)$ a density of states, but there is a difficulty with this interpretation.  
If $\rho(E)$ were a sum of delta functions with integer coefficients,
say $\rho(E)=\sum_i n_i \delta(E-E_i)$, we would interpret the Schwarzian theory as a quantum system 
with energy levels $E_i$, of multiplicity $n_i$.  Instead $\rho(E)$ is a smooth function of $E$.  One might think that this means that the Schwarzian theory 
is a quantum system with a continuous energy spectrum, but a moment's reflection shows
that that interpretation is not viable.  In general, if $\H$ is a Hilbert space and $H$ is a 
Hamiltonian operator acting on $\H$ with continuous spectrum, then $\Tr_\H\,\exp(-\beta H)$ is divergent
for any $\beta$.  (If $H$ has continuous spectrum in an 
interval $E_0\leq E\leq E_1$, then it has infinitely many energy levels in any subinterval of  the interval $[E_0,E_1]$.  The sum over
this infinity causes $\Tr_\H\,\exp(-\beta H)$ to diverge.)

In general, in physics, whenever one talks about a continuous density of states per unit energy, one always really has in mind 
the density of states per unit energy per unit ``$x$,'' where
$x$ is some other variable.  For example, a free particle in a box has a discrete spectrum of energy levels, but 
a free particle in infinite volume has a continuous density of states per unit
energy per unit volume.  Thus, by itself the partition function $Z(\beta)$ of the Schwarzian theory is not $\Tr_\H \, \exp(-\beta H)$ for any
quantum system, though it might be a factor in the partition function of a quantum system (such as the SYK model) in a suitable limit
(such as the large $N$ limit).

Technically, the fact that the Schwarzian path integral does not have a Hilbert space interpretation as  $\mathrm{Tr}_\H\,\exp(-\beta H)$
results from the fact that in defining this path integral, one divides by $SL(2,\R)$, a step that would not be available if $\tau$ is taken
to parametrize  a real line rather than a circle.

\section{Generalizations}\label{sthree}
The Schwarzian theory has a family of generalizations, involving integrals over coadjoint orbits of Lie groups that include the Virasoro group. There are two facts that make it possible to get a one-loop exact theory this way: {\it (i)} these orbits are always symplectic manifolds, and {\it (ii)} they always have a $U(1)$ time translation generator associated to $L_0$ that we can use as an action. It is interesting that some of the theories one defines this way have already been shown to arise from generalizations of the SYK model.

\subsection{More Details on the Virasoro Case}
To begin, we will describe how one arrives at the ordinary Schwarzian theory using this perspective. Here the relevant integration space, $\text{diff}(S^1)/SL(2,\R )$, is a coadjoint orbit of the Virasoro group.  Virasoro is a central extension of $\text{diff}\,S^1$, and the central extension will play an essential role.

To understand the Virasoro coadjoint representation, it is helpful to briefly review the more familiar adjoint representation. An element of the adjoint representation is a linear combination $\sum_n v_n L_n$ of the Virasoro generators $L_n$, plus a multiple $a I$ of the central element $I$. (The eigenvalue of $I$ in a given representation is usually called $c$, the central charge.)   It is convenient to Fourier transform and consider $v(\tau) = \sum_n e^{in\tau}v_n$ and $L(\tau) = \sum_ne^{-in\tau}L_n$.   The Virasoro algebra, including the central extension, is \be
[L_m,L_n] = (m-n)L_{m+n} + \frac{I}{12}m^3\delta_{m,-n}, \hspace{20pt} [L_n,I] =0.
\ee
Then an element of the adjoint representation  is given by a pair $(v(\tau),a)$, which represents a Lie algebra element $\int \frac{d\tau}{2\pi} v(\tau)L(\tau) + aI $. 
An orbit in the  Lie algebra is obtained by conjugating this
element by a group element $g = \exp\left[i\int \frac{d\tau'}{2\pi}w(\tau')L(\tau')\right]$. This results in 
\be\label{adjointRep}
\text{Ad}_{\phi^{-1}}(v(\tau),a)= \left(\frac{v(\phi(\tau))}{\phi'(\tau)},a + \frac{1}{12}\int \frac{d\tau}{2\pi}\frac{v(\phi(\tau))}{\phi'(\tau)}\sch(\phi,\tau)\right).
\ee
Here $\phi(\tau)$ is related to $w(\tau)$ by $\phi(\tau) = e^{-w(\tau)\partial_\tau}\tau e^{w(\tau)\partial_\tau}$. In other words, $\phi^{-1}$ is the diffeomorphism that we get by exponentiating the vector field $w(\tau)$. We can understand the transformation of $v$ in a simple way as that of a vector field transforming under a diffeomorphism given by $\phi^{-1}$.

Now we would like to understand the coadjoint representation. This is simply the dual space of the adjoint representation. It can be parametrized as\footnote{The reason that we call the second parameter here $c$ is that the Virasoro
representation obtained by quantizing this orbit does have central charge $c$.  In other words, upon quantization, $c$
becomes the eigenvalue of the central element $I$ of the Virasoro algebra.} $(b(\tau),c)$, with the pairing
\be\label{pairing}
\langle (b(\tau),c),(v(\tau),a)\rangle = \int_0^{2\pi} \frac{d\tau}{2\pi}b(\tau)v(\tau) + ca.
\ee
The transformation of $(b(\tau),c)$ under the Virasoro group is determined by requiring that this pairing should be invariant when we transform both $(b(\tau),c)$ and $(v(\tau),a)$. It is easy to work out the necessary transformation using (\ref{adjointRep}). It turns out that the action preserves $c$, which is thus a constant that characterizes
the orbit (and after quantization, the representation).
This of course reflects the fact that $I$ is central and is dual to the fact that in the adjoint representation, the transformation of $v$ does not depend on $a$. The action on $b$ is 
\be\label{coadjointRep}
b_\phi\equiv \text{Ad}^*_{\phi^{-1}}\left(b(\tau)\right)= \phi'(\tau)^2b(\phi(\tau)) - \frac{c}{12}\sch(\phi,\tau).
\ee
We can interpret this formula by saying that $b$ transforms like (minus two times) a stress tensor in a 2d CFT.

Eq.~(\ref{coadjointRep}) defines the coadjoint representation. Coadjoint orbits are given by picking an initial $b(\tau)$ and then varying $\phi$. Since some $\phi$ configurations lead to the same coadjoint vector, the orbit is identified with a quotient of $\text{diff}(S^1)$, where the subgroup that we quotient by is the subgroup of $\text{diff}(S^1)$ that leaves the coadjoint vector unchanged. For example, if we choose an initial $b$ that is some generic constant value, then $\text{Ad}^*_{\phi^{-1}}(b)$ will only be invariant under constant shifts of $\phi$, and we get an orbit $\text{diff}(S^1)/U(1)$. However, if we choose the special constant value $b(\tau) = -\frac{c}{24}$, then we can write
\be\label{online}
\text{Ad}_{\phi^{-1}}^*\left(-\frac{c}{24}\right) = -\frac{c}{12}\sch\left(\tan\frac{\phi}{2},\tau\right).
\ee
It is clear that this is invariant under $SL(2,\R)$ transformations of $f \equiv \tan\frac{\phi}{2}$, so the orbit in this case is $\text{diff}(S^1)/SL(2,\R)$.   (This value of $b$ corresponds after quantization to the Virasoro representation that
comprises the identity operator and its descendants.)

Now, to relate this orbit to the Schwarzian theory, we will quote two general facts.  First, the orbit is a symplectic manifold. The symplectic form given by the pairing $\omega = \langle b_\phi,[v,v']\rangle$ where $v,v'$ are elements of the adjoint representation, which parametrize the tangent space to the coadjoint orbit. We can also parametrize this tangent space in terms of infinitesimal changes in $\phi$. The relation between these is given by $v(\tau) = d\phi^{-1}(\tau) = -\frac{d\phi(\tau)}{\phi'(\tau)}.$ 
It is easy to check that this identification and the Virasoro algebra leads to the expression for the symplectic form given in (\ref{omega}). Second, the symplectic generator of the group action associated to a given Lie algebra element is simply given by taking the pairing of $b_\phi$ with that element. If we take the Lie algebra element to be $L_0$, which generates $\tau \rightarrow \tau + \epsilon$, then we get the generator by taking the pairing of $b_\phi$ with a constant vector field. Using (\ref{pairing}) and (\ref{online}) this immediately gives the Schwarzian action discussed in the previous section.

Of course, we can also consider a more generic orbit, such as the orbit of some constant $b = b_0$. This does not arise in the SYK model, but it results in a perfectly reasonable one-loop exact path integral. In order for the theory to be stable for small fluctuations about the saddle point $\phi = \tau$, we require that $b_0 > -\frac{c}{24}$. Then the orbit is $\text{diff}(S^1)/U(1)$, and the partition function is
\be
Z(g) = \frac{\#}{g}\exp\left(\frac{-24b_0}{c}\cdot\frac{\pi}{g^2}\right).
\ee
Note that the one-loop determinant in this case is proportional to $\frac{1}{g}$ instead of $\frac{1}{g^3}$, because we have only one zero mode to gauge fix in the integral over $\phi$.

\subsection{Virasoro-Kac-Moody}
One simple generalization is to extend the Virasoro algebra to a Virasoro-Kac-Moody algebra with group $G$, and repeat the above steps. One finds an action that is the sum of the Schwarzian action plus the quantum mechanics for a particle moving on the group manifold, $L = \text{tr}\left[(g^{-1}\partial_\tau g)^2\right]$. The partition function for such a particle was already known to be one-loop exact \cite{0305-4470-22-13-024}. One expects this theory to be relevant at low-energies for generalizations of SYK with global symmetry \cite{Gu:2016oyy,Gross:2016kjj,Witten:2016iux,Klebanov:2016xxf}.

\subsection{Super-Virasoro}
We can get an interesting generalization by considering super-Virasoro algebras. The coadjoint representations of the $\mathcal{N} = 1$ and $\mathcal{N} = 2$ algebras were considered in \cite{Bakas:1988mq,Delius:1990pt,0253-6102-16-3-295}. We can understand the transformation of the coadjoint vector in a simple way as the transformation of the super stress tensor under superconformal transformations. For the application to SYK, we are interested in the orbit that corresponds to a super  stress tensor that vanishes on the line (where $f = \tan\frac{\phi}{2} = \tau$), and is invariant under global superconformal transformations. Taking this orbit, we can get an action for a one-loop exact theory by pairing the coadjoint vector with $L_0$. Concretely, the resulting action is the integral over super space of the super Schwarzian derivative. Such actions were previously argued to arise as the low-energy theories of $\mathcal{N} = 1$ and $\mathcal{N} = 2$ supersymmetric SYK models \cite{Fu:2016vas}.

The point of relating these theories to a coadjoint orbit is that it follows that the partition functions are one-loop exact.\footnote{This relies on some small extensions of the bosonic facts discussed above. First, it is easy to check directly (or see \cite{Kostant:1975qe}) that coadjoint orbits of super Lie algebras are symplectic supermanifolds. And second, the Duistermaat-Heckman formula generalizes to integrals over symplectic supermanifolds, as we show in Appendix \ref{appendixDH}.} We will make a few comments about both the $\mathcal{N} = 1$ and $\mathcal{N} = 2$ cases. For $\mathcal{N} = 1$, the super Schwarzian theory includes a bosonic field $f = \tan\frac{\phi}{2}$ and an antiperiodic Grassmann field $\eta$. The action is invariant under the global superconformal transformations, $OSp(1|2)$, which we interpret as a gauge symmetry, just as in the bosonic case. In other words, we integrate over all superconformal transformations modulo $OSp(1|2)$.\footnote{The theory also has a physical supersymmetry that is broken by the thermal (antiperiodic) boundary conditions for $\eta$. Note that $OSp(1|2)$ is not broken by this condition.} The classical term in the partition function is the same as in the bosonic model, $I = -\frac{\pi}{g^2}$. The one-loop term is proportional to an expression involving the number of bosonic and fermionic zero modes that we are quotienting by: $g^{\#_{f} - \#_{b}}$. The group $OSp(1|2)$ has three bosonic and two fermionic generators, so we get a one-loop factor $\frac{1}{g}$. Translating $g$ into dependence on the inverse temperature $\beta$ and a parameter $C = \beta/(2\pi g^2)$ proportional to the specific heat, we find
\be
Z(\beta) \propto \frac{1}{\beta^{1/2}}\exp\left(\frac{2\pi^2 C}{\beta}\right) \hspace{20pt} \implies\hspace{20pt} \rho(E) \propto \frac{\cosh(2\pi\sqrt{2CE})}{E^{1/2}}.
\ee
Notice that in this case, the density of states has a square-root growth at low energies.

In the $\mathcal{N} = 2$ case, we integrate over the usual bosonic field $f = \tan\frac{\phi}{2}$, two antiperiodic fermions $\eta, \bar{\eta}$, and a compact scalar $\sigma$, which is defined up to $\sigma \sim \sigma + 2\pi n \hat{q}$, where $\hat{q}$ is a parameter of the model. In the application to SYK, $\hat{q}$ is an odd integer that determines the number of fermions appearing in the supercharge \cite{Fu:2016vas}. The global super-conformal group is $SU(1,1|1)=OSp(2|2)$, which has four bosonic and four fermionic generators, giving a one-loop determinant that is independent of $g$. The main new feature in the $\mathcal{N} = 2$ case is that we have to sum over a family of saddles, where $\sigma$ winds $n$ times around the thermal circle \cite{Fu:2016vas}. We will use formulas for the super-Schwarzian action given in \cite{Fu:2016vas} and refer the reader there for details. The purely bosonic part of the action is
\be
I_b = \frac{1}{g^2}\int_0^{2\pi} d\tau \left[-\sch(\tan\frac{\phi}{2},\tau) +2(\partial_\tau\sigma)^2\right], \hspace{20pt} g^2 = \frac{\beta}{2\pi C},
\ee
Up to $SU(1,1|1)$ gauge transformations, the saddles are $\phi = \tau$ and $\sigma = n\hat{q}\tau$. The action for such a saddle is $I_b = -\frac{\pi}{g^2}(1 - 4n^2\hat{q}^2)$, and the bosonic one-loop determinant is independent of $n$. The fermionic part of the action is somewhat more complicated, but we will only need the quadratic part to compute the one-loop determinant. Expanding eqn. (5.32) from \cite{Fu:2016vas} about the saddle just described, we find
\begin{align}
I_{f,\,quad} &= \frac{1}{g^2}\int_0^{2\pi}d\tau \left[\eta \partial_\tau \bar{\eta} - 4(\partial_\tau\eta)(\partial_\tau - in\hat{q})^2\bar{\eta}\right]\\
&=\frac{2\pi i}{g^2}\sum_{m} (4m^2-1)(m-n\hat{q})\eta_{m-n\hat{q}}\bar{\eta}_{-m+n\hat{q}}.
\end{align}
In the second line we have expanded $\eta$ in modes, $\eta(\tau) = \sum_{p} \eta_p e^{ip\tau}$. We should impose antiperiodic (thermal) boundary conditions, so $m$ and $p$ are half-integers. The two zero modes for each fermion are $m = \pm\frac{1}{2}$. We are interested in the dependence of the determinant on the saddle point label $n$. To compute this, we can regularize by dividing by the determinant with $n = 0$. Taking the product over all non-zero modes, we get
\be
\prod_{m =...-\frac{5}{2},-\frac{3}{2},\frac{3}{2},\frac{5}{2},...}\frac{m-n\hat{q}}{m} = \prod_{m = \frac{3}{2},\frac{5}{2},...}\left(1 - \frac{n^2\hat{q}^2}{m^2}\right) = \frac{\cos(\pi\hat{q}n)}{1 - 4 \hat{q}^2n^2}
.
\ee
So the contribution from a given saddle is proportional to 
\be
Z_n(\beta) = \frac{\cos(\pi \hat{q}n)}{1 - 4 \hat{q}^2n^2}\exp\left[\frac{2\pi^2 C}{\beta}\left(1 - 4 n^2\hat{q}^2\right)\right].
\ee
Since $\hat{q}$ is odd, the factor of $\cos(\pi\hat{q}n)$ can be written for integer $n$ as $(-1)^n$, but we will leave it in this form for the moment, because we will consider non-integer $n$ below.

To compute the contribution of such a saddle to the density of states, we use the integral
\be
\int_0^\infty dE\left[\delta(E) + \sqrt{\frac{a}{E}}I_1(2\sqrt{aE})\right]e^{-\beta E} = e^{\frac{a}{\beta}}.
\ee
This is true for both positive and negative $a$, provided that the square root is defined the same way inside and outside the Bessel function. It follows that a single saddle point gives a contribution to the density of states proportional to
\be
\rho_n(E)= \frac{\cos(\pi \hat{q}n)}{1 - 4\hat{q}^2n^2}\left[\delta(E) + \sqrt{\frac{a_n}{E}}I_1(2\sqrt{a_nE})\right]\hspace{20pt} a_n\equiv 2\pi^2 C(1 - 4n^2\hat{q}^2).
\ee

We would like to determine which sum over saddles is appropriate for the SYK theory. This can be done by requiring consistency when the system is coupled to a chemical potential. Concretely, we imagine adding a factor $e^{i\alpha Q}$ in the thermal trace, where $Q$ is the $U(1)$ charge, normalized so that the original SYK fermions carry charge one. Inserting this factor is the same as doing the path integral with boundary conditions twisted by the $U(1)$ rotation. The action of this twist on the fields of the super-Schwarzian theory is
\be
\sigma \rightarrow \sigma+\hat{q}\alpha, \hspace{20pt} \eta \rightarrow e^{-i\hat{q}\alpha}\eta, \hspace{20pt} \bar{\eta}\rightarrow e^{i\hat{q}\alpha}\bar{\eta}.
\ee
This has precisely the same effect as adding to the integer saddle-point parameter $n$ a fractional part $\frac{\alpha}{2\pi}$, so the contribution of a given saddle gets modified as $Z_n\rightarrow Z_{n + \frac{\alpha}{2\pi}}$.

\begin{figure}
\begin{center}
\includegraphics[width=.24\textwidth]{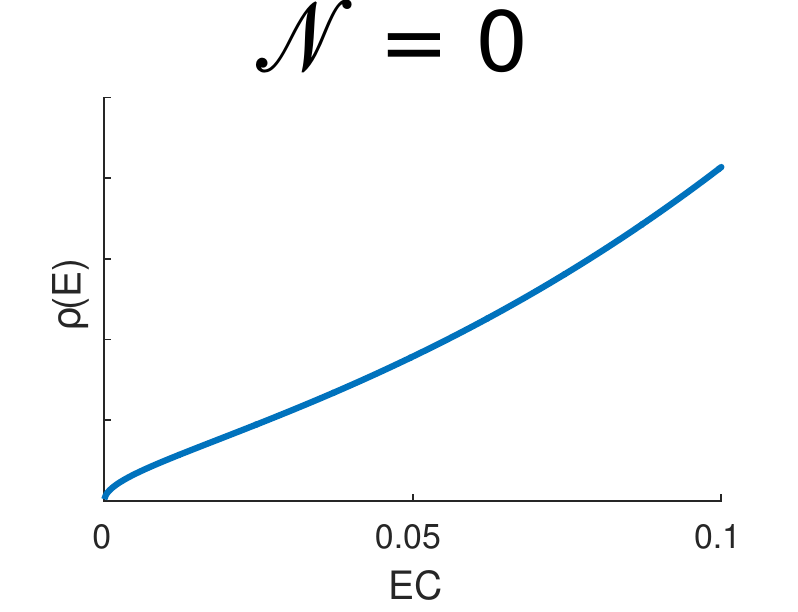}
\includegraphics[width=.24\textwidth]{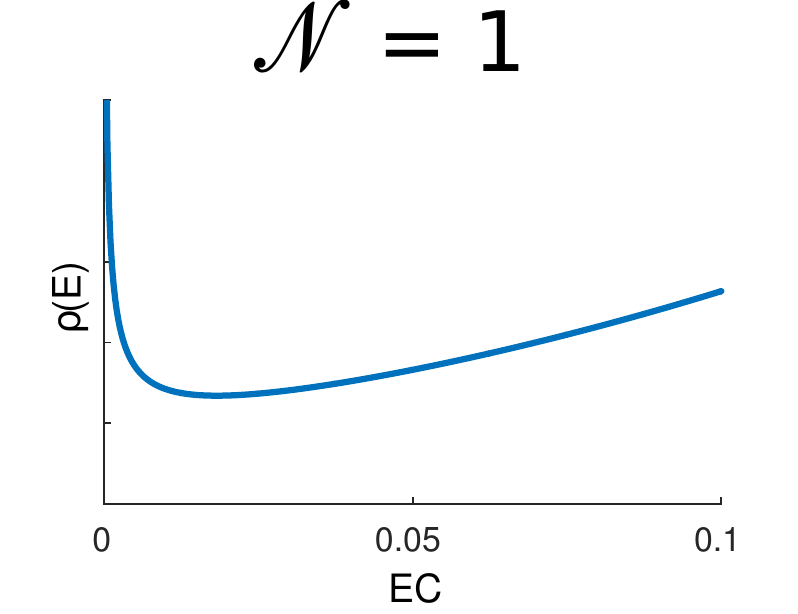}
\includegraphics[width=.24\textwidth]{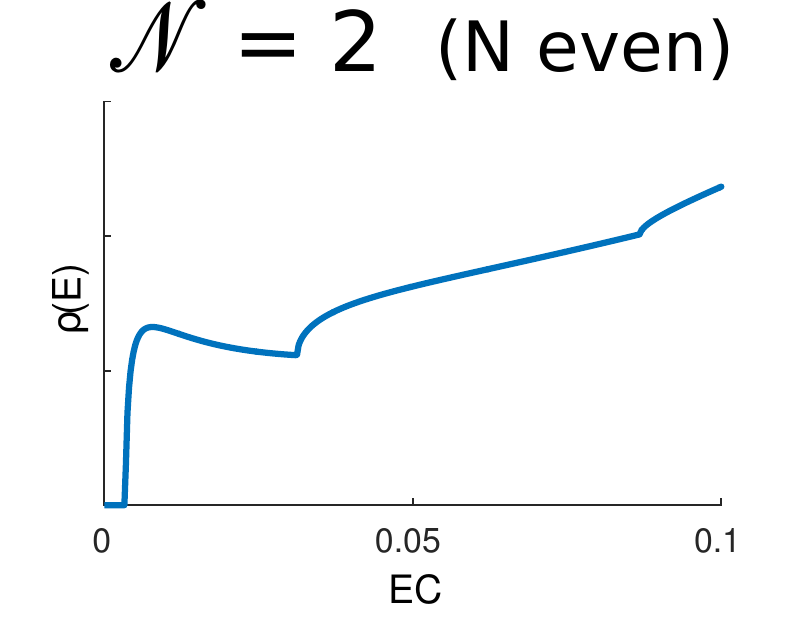}
\includegraphics[width=.24\textwidth]{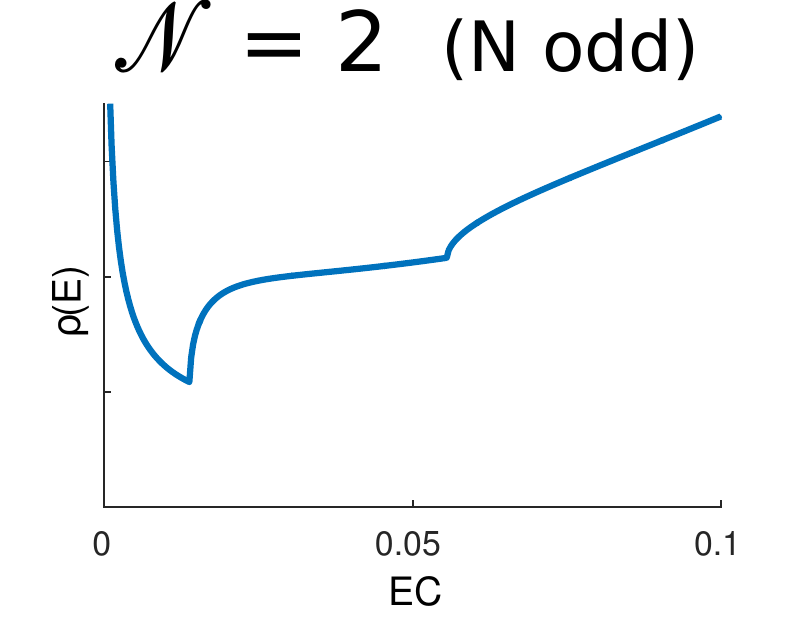}
\includegraphics[width=.24\textwidth]{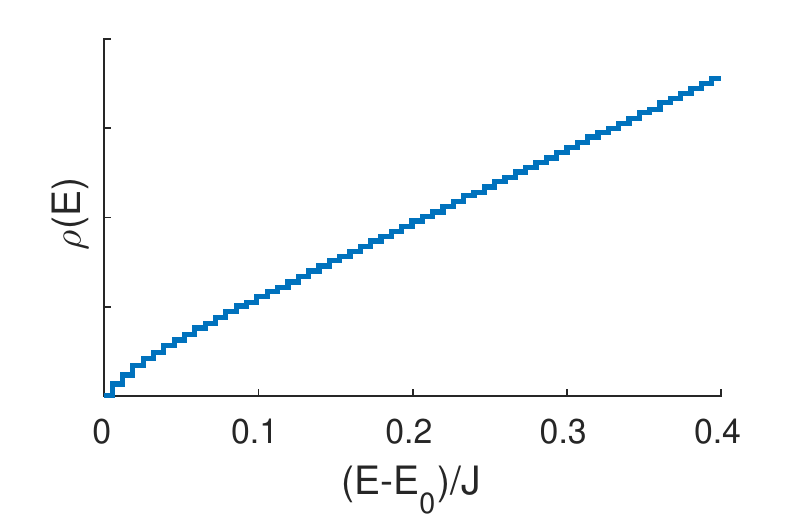}
\includegraphics[width=.24\textwidth]{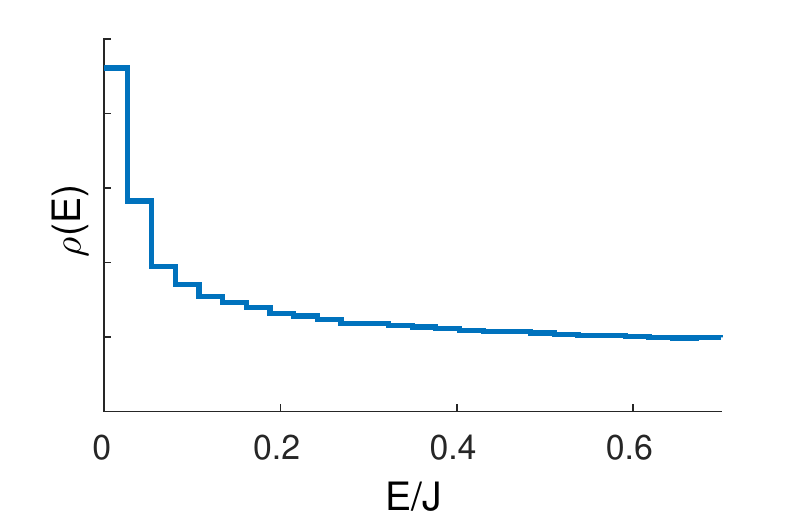}
\includegraphics[width=.24\textwidth]{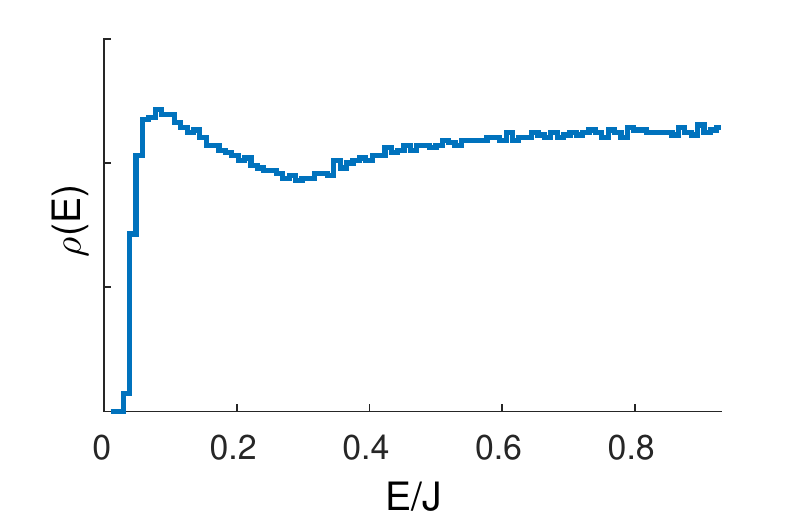}
\includegraphics[width=.24\textwidth]{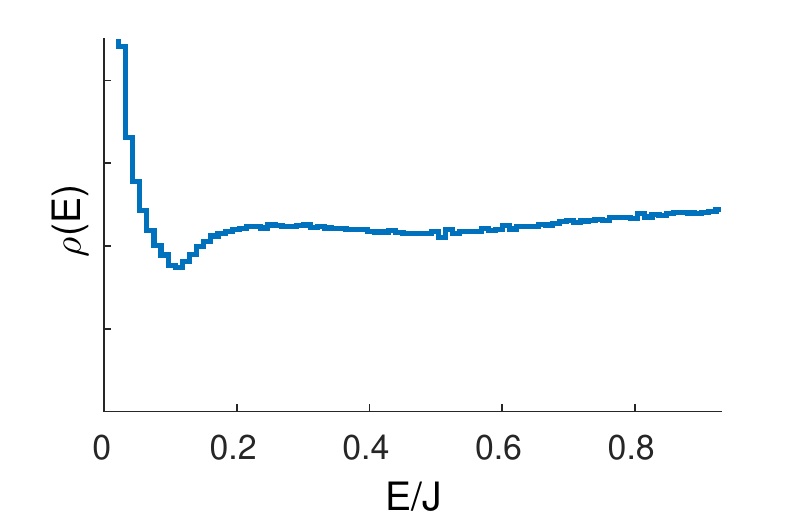}
\caption{{\bf Top:} Exact density of states of the Schwarzian theory near the ground state, for different amounts of supersymmetry. For the two $\mathcal{N} =2 $ cases we chose $\hat{q} = 3$ and we omitted the $\delta(E)$ contribution to the spectrum. {\bf Bottom:} The numerical density of states from exact diagonalization of (super) SYK with a four-fermion Hamiltonian. The Hilbert space dimensions used are respectively $2^{16},2^{16},2^{18},2^{19}$. The $\mathcal{N} = 0$ case is an average using data from \cite{Cotler:2016fpe}. The other curves are for a single realization.}\label{rhoofeplots}
\end{center}
\end{figure}

This observation allows us to detemine how to sum over saddles, as follows. We have to separately consider two cases, depending on whether the number of complex SYK fermions, $N$, is even or odd. In the case of even $N$, the possible values of the $U(1)$ charge are integers, so if we set $\alpha = 2\pi$, then $e^{i\alpha Q} = 1$ and the total partition function must be the same. Increasing $\alpha$ from zero to $2\pi$ has the effect of mapping $Z_{n}\rightarrow Z_{n+1}$, so to have a total partition function that is invariant, we need to sum over all values of $n$ with the same coefficient. On the other hand, when $N$ is odd, the possible charges are half-integer, so when we set $\alpha = 2\pi$ we have $e^{i\alpha Q} = -1$ and the total partition function should change by a sign. We can accomplish this by summing over even $n$ saddles with coefficient one, and odd $n$ saddles with coefficient minus one. This gives the density of states
\be
\rho_{\text{even}}(E) =\sum_{n = -\infty}^\infty \rho_n(E) \hspace{40pt} \rho_{\text{odd}}(E) = \sum_{n=-\infty}^\infty (-1)^n\rho_n(E),
\ee
where the subscript indicates whether $N$ is even or odd. These expression for $\rho$ are plotted in figure \ref{rhoofeplots}. To explain the rather strange form of the resulting curves, it is helpful to think about applying Poisson summation to these formulas. In both the even and odd cases, we can write the answer as
\begin{align}\label{justified}
\rho(E) \propto \sum_m \rho(E,m), \hspace{20pt} \rho(E,m) \equiv \int dn e^{2\pi i m n}\rho_n(E),
\end{align}
The only difference is that in the even case we sum over integer $m$, and in the odd case, we sum over half-integer values of $m$ (and not integer values). In the even case, this is a straightforward application of the Poisson summation formula. In the odd case, it requires an extra step where we apply Poisson summation separately to the even $n$ and odd $n$ sums, only combining terms at the end. This gives
\be\label{rhoodd}
\rho_{\text{odd}}(E) = \sum_{2m\in \Z} \rho_{\text{odd}}(E,m)
\ee
where
\begin{align}
\rho_{\text{odd}}(E,m) = \int dn e^{2\pi i m n} \left[\rho_n(E) - \rho_{n+1}(E)\right] = (1 - e^{-2\pi i m n})\rho(E,m).
\end{align}
In principle, the sum in (\ref{rhoodd}) involves both half-integer and integer values of $m$, because $m$ is the Poisson-summation dual of a variable $n$ that is being summed over even values. However, we see that the summand vanishes for integer values of $m$, so in practice we can sum only over half-integer $m$.

The point of writing the density of states as in (\ref{justified}) is that we can think of the fourier transform as an integral over imaginary values of the chemical potential, which has the effect of selecting the contribution of states of charge $m$. So $\rho(E,m)$ is the density of states in charge sector $m$. 

It is interesting to consider the range of energies that contribute in the various charge sectors. It is straightforward to check that the $\delta(E)$ contribution is present only for charges that satisfy $|m| < \frac{\hat{q}}{2}$. For the continuum part of the spectrum, we have to consider the fourier transform of the expression including the Bessel function. Despite appearances, the integrand is an entire function of $n$. In fact, this is also true for each term if we write the cosine as a sum of two exponentials (the naive pole at $1 - 4\hat{q}^2n^2 = 0$ is canceled by a zero of the Bessel function expression). Using the asymptotic behavior of the Bessel function at large $n$, we find that the contour can be closed either in the upper half plane or in the lower half plane (for $|m|<\frac{q}{2}$ we close in different directions for the two terms in the cosine) unless $E > E_0(m)$, with
\be
E_0(m) = \frac{1}{2C}\left(\frac{|m|}{2\hat{q}}-\frac{1}{4}\right)^2.
\ee
This is the lowest energy of the continuum part of the spectrum in charge sector $m$. The fact that different charge sectors start contributing rather sharply at different energies explains the odd shape of $\rho(E)$ in figure \ref{rhoofeplots}. Remembering that $\hat{q}$ is odd, we can see that  if $N$ is even (so that $m$ is an integer), we have that $E_0(m)$ is positive for all values of $m$. This means that there is a gap above the degenerate ground states described by the $\delta(E)$ term. By contrast, in the case of odd $N$, we have sectors of charge $m = \pm\frac{\hat{q}}{2}$ that are gapless.

The qualitative features of the low energy spectrum derived in the $\mathcal{N} = 0$, $\mathcal{N} = 1$ and $\mathcal{N} = 2$ theories all agree quite well with exact diagonalziation numerics for the SYK model, again at low energies. This is particularly impressive in the $\mathcal{N} = 2$ case, where the story is fairly complicated. We will not attempt a quantitative fit of these curves, but we compare informally in figure \ref{rhoofeplots}.

\section*{Acknowledgements}
We are very grateful to Wenbo Fu, Guy Gur-Ari, Juan Maldacena, Greg Moore, Nikita Nekrasov, Gabor Sarosi, Steve Shenker, and David Simmons-Duffin for discussions. D.S. is supported by the Simons Foundation grant 385600.  E.W. is supported in part by NSF Grant PHY-1606531.

\appendix

\section{Duistermaat-Heckman for Supermanifolds}\label{appendixDH}
In this appendix, we give a physics proof of a Duistermaat-Heckman formula for supermanifolds. The formula
says that on a symplectic supermanifold, the integral of the exponential of a generator of a $U(1)$ symmetry is one-loop exact.   This is a small generalization of the original DH formula, which was for integrals over bosonic symplectic manifolds. In the proof of the bosonic version of the formula
that was  sketched in the introduction, one introduces fermionic partners for the original purely bosonic integration variables. In the super case, we will introduce fermionic partners for the bosons, and bosonic partners for the fermions.

Let us make this more explicit. The starting point is a supermanifold $M$ with bosonic coordinates $t^1\dots t^p$ and fermionic coordinates $\theta^1\dots \theta^q$. We will use $x^A$ as a general coordinate that runs over both fermionic and bosonic values. For each bosonic coordinate, we introduce a fermionic partner that we call $dt^a$, and for each fermionic coordinate we introduce a bosonic partner that we call $d\theta^r$ (in the introduction, we called the fermions $\psi^a$ instead of $dt^a$). The reason that we introduce these coordinates is to write the symplectic measure as an integral. The reason that we refer to them as $dt,d\theta$ is that one can think about differential forms on $M$ simply as functions of this enlarged set of coordinates.\footnote{The space of $t,\theta,dt,d\theta$ is sometimes called $\Pi TM$, meaning a statistics-reversed version of the tangent bundle to the original manifold $M$. See e.g. \cite{Witten:2012bg} for details.} For example, the symplectic two-form is a function of the type
\be\label{fundef}
\omega = dt^a \omega_{ab}dt^b + dt^a\omega_{ar}d\theta^r + d\theta^r\omega_{ra} dt^a+ d\theta^r \omega_{rs}d\theta^s = dx^A\omega_{AB}dx^B.
\ee
Here, the supermatrix of components $\omega_{AB}$ are functions of the coordinates $t,\theta$ but not of their partners $dt,d\theta$. We take $\omega$ to be even-valued, which requires that $\omega_{ab},\omega_{rs}$ are even, while $\omega_{ar}$ is odd. The meaning of the statement that we have a symplectic manifold is that {\it (i)} this form is nondegenerate, in the sense that the supermatrix $\omega_{AB}$ is invertible at each point on our manifold, and {\it (ii)} $d\omega = 0$ where we define
\be
d = dt^a\frac{\partial}{\partial t^a} + d\theta^r \frac{\partial}{\partial \theta^r} = dx^A\frac{\partial}{\partial x^A}.
\ee
Note that $d$ is an ordinary differential operator acting in the space of functions of $t,\theta,dt,d\theta$.

Now, we assume that there is a function $H$ that generates a $U(1)$ symmetry of the manifold via the Hamiltonian flow
\begin{align}\label{flowH}
\delta x^A  = v^A = -\partial_B H\omega^{BA}
\end{align}
where $\omega^{AB}$ is the inverse of the supermatrix $\omega_{AB}$. Any function $H$ generates a flow that preserves $\omega$ and can be understood as a symmetry of the manifold. However, it is crucial that we are assuming the flow generated by $H$ is a $U(1)$ symmetry, so that all orbits close after the same amount of time. Now, we consider the integral
\be\label{integr}
Z = \int D(x^A, d x^A)\exp
\left(\frac{1}{2}\omega +H\right)
,
\ee
where the integral runs over all of the variables $t,\theta,dt,d\theta$ with various indices and  $\omega$ is the function defined in eqn. (\ref{fundef}).  Just as  in the introduction, there
is a natural measure $D(x^A,dx^A)$ because the variables $dx^A$ transform the same way as $x^A$ but with opposite statistics.  In eqn. (\ref{integr}), the integral over the even variables $t$
and $d\theta$ is an ordinary integral, and the integral over the odd variables
$\theta$ and $dt$  is defined by the standard Berezin rules.  Since the exponent in (\ref{integr})  is quadratic in $d\theta$ and $d t$ ($\omega$ is homogeneous and
quadratic in those variables and $H$ does not depend on them),
the integral in (\ref{integr}) is Gaussian in the $d\theta$ and $dt$ variables. We can imagine doing these integrals first. The effect of doing these integrals is to give
 the symplectic 
measure for the remaining integrals over $t,\theta$.  This measure 
is $d^px d^q\theta \sqrt{\Ber(\omega_{AB})}$.  Here $\Ber$ is the Berezinian, the superanalog
of the determinant.  So the integral in (\ref{integr}) is the integral over the original supermanifold of $\exp(H)$.

If $M$ is an ordinary (bosonic) symplectic manifold, then instead of $\sqrt{\det \omega}$ we have the more natural $\Pf(\omega)$, but in the superworld there is no analog
of the Pfaffian and we have to use the square root of the Berezinian. This is related to the following circumstance.  If $\omega$ is an ordinary symplectic form on a manifold (or just on
a vector space) then it determines an orientation.  This orientation determines a natural sign of the square root $\sqrt{\det\omega}$, and the square root with that sign is called the Pfaffian.
On a symplectic supermanifold,  the even tangent bundle\footnote{Technically, the even and odd tangent bundles of $M$ are naturally defined over
the reduced space of $M$, and not over $M$ itself.  For the purposes of the topological point that we are explaining here, this difference is immaterial.} of $M$ still has a natural orientation,
but the odd tangent bundle of $M$ does not.  Accordingly, there is no natural sign for the fermionic measure $d^q\theta$.  The upshot
is that although there is a natural symplectic measure that we can write informally as  $d^pxd^q\theta \sqrt{\text{Ber}(\omega)}$, the factors $d^px d^q\theta$ and $\sqrt{\text{Ber}(\omega)}$
do not separately have naturally-defined signs.  (Going back to eqn. (\ref{integr}), the overall measure $D(x^A,dx^A)$ is completely well-defined, but if we write it as the product of a
measure for $x,\theta$ and a measure for $dx,d\theta$, then both factors separately have sign problems.)

Now, we define a supersymmetry by
\be\label{superS}
Qx^A = dx^A, \hspace{20pt} Q(dx^A) = v^A,
\ee
where the variation $v^A$ is the flow generated by $H$, see (\ref{flowH}). It is straightforward to check using the fact that $d\omega = 0$ that the variation of the action vanishes, $Q(\frac{1}{2}\omega + H) = 0$. One can also show that $Q^2 x^A = v^A$ and $Q^2(dx^A) = d(v^A)$ so that $Q^2$ is simply the generator of the $U(1)$ flow associated to $H$. The final preliminary step is to choose a $U(1)$ invariant metric $g_{AB}$ (one can take any metric and average it over $U(1)$ to make it invariant; this is where it is important $H$ generates a compact $U(1)$ symmetry rather than a generic symplectomorphism).

To make the localization argument, we now define a Grassman-odd function
\be
V = g_{AB}v^A dx^B.
\ee
From the $U(1)$ invariance of the metric and the fact that $Q^2$ is the $U(1)$ generator, it follows  that $Q^2V = 0$. We can therefore localize by adding $s QV$ to the action with a large coefficient $s$. This does not change the integral; indeed, the integral of a $U(1)$-invariant $Q$-exact
function vanishes, and the formula
\be
e^{\frac{1}{2}\omega +H - sQV} = e^{\frac{1}{2}\omega+H} + Q\left[e^{\frac{1}{2}\omega+H}\left(-sV + \frac{s^2}{2}VQV + ...\right)\right],
\ee
shows that all terms that depend on $s$ are $Q$-exact.

The term that we added does not change the integral, but it does change the integrand, and it has the effect of localizing the integral to the critical points of $H$. To see this, we write
\be\label{QV}
QV =  dx^C\frac{\partial g_{AB}}{\partial x^C}v^Adx^B + dx^C\frac{\partial v^A}{\partial x^C}dx^B g_{AB}+ v^Bg_{AB}v^A.
\ee
When the localization parameter $s$ is large, we are localized to the region where $v^A$ and $dx^A$ are zero. This will be a location where the $\theta$ variables vanish, and the $t$ variables take some particular value, say $t_*$. We can then expand about this point as
\be
v^A = y^B \partial_B v^A(t_*), \hspace{20pt} y^a = t^a - t^a_*, \hspace{20pt} y^r = \theta^r.
\ee
We then see that the first term in (\ref{QV}) is cubic in the small deviation from the critical point and can be ignored relative to the second and third terms, which are quadratic. The quadratic action is
\be
QV_{\text{gaussian}} = \left[g_{CB}\partial_Av^C\right]_{\theta = 0, t = t_*}dx^Adx^B + \left[g_{DC}\partial_A v^C\partial_B v^D\right]_{\theta = 0,t = t_*}y^Ay^B.
\ee
Notice that the quantities in brackets are bosonic, depending only on $t_*$, so the quadratic terms only couple variables of the same fermionic parity. So we have separate Gaussian integrals over the $t,\theta,dt,d\theta$ variables. After doing these integrals, cancelling factors of the determinant of the metric and the localization parameter $s$, and using $\partial_A v^B(t_*) = -\partial_A(\omega^{CB}\partial_C H)|_{t_*} = -\omega^{CB}(t_*)\partial_A\partial_C H(t_*)$, we find that the contribution from the critical point $t_*$ is (up to measure factors of $2\pi$) 
\be
Z_{t_*} = e^{H(t_*)}\frac{\sqrt{\det\left[\partial_r\partial_s H\right]}}{\sqrt{\det \omega_{rs}}}\frac{\sqrt{\det \omega_{ab}}}{\sqrt{\det\left[\partial_a\partial_b H\right]}}\Bigg|_{\theta = 0,t=t_*} = e^{H(t_*)}\frac{\sqrt{\text{Ber}\,\omega_{AB}}}{\sqrt{\text{Ber}\left[\partial_A\partial_B H\right]}}\Bigg|_{\theta = 0,t = t_*}.
\ee
Here, as above, we use $r,s$ for fermionic indices, and $a,b$ for bosonic indices. Notice that mixed derivatives of $H$ and mixed components of $\omega$ vanish at the critical point, because they would have to be odd and therefore proportional to odd powers of $\theta$.  As we have explained above, there is no natural sign of the square root $\sqrt{\text{Ber}\,\omega_{AB}}$,
but $\sqrt{\text{Ber}\,\partial_A\partial_B H }$ has precisely the same problem and the ratio of the two square roots has a well-defined sign.  (Concretely, the sign of
 $\sqrt{\text{Ber}\,\omega_{AB}}$ and the sign of 
 $\sqrt{\text{Ber}\,\partial_A\partial_B H }$ depend on an orientation of the odd tangent bundle of $M$, but this dependence is absent in the ratio.)

If the function $H$ has more than one critical point (or equivalently if the $U(1)$ action on $M$ has more than one fixed point),
then  we should sum over these points to get the full answer for the integral.  As in the case of the bosonic Duistermaat-Heckman formula, if the critical points
form a moduli space $\mathcal M$ of positive dimension, one can express the original integral over $M$ as an integral over $\M$.   (For instance -- although this example involves
a further generalization to a nonabelian analog of Duistermaat-Heckman -- in the case of two-dimensional Yang-Mills theory, one can take $M$ to be the space of all gauge fields on
an oriented  two-manifold and then $\M$ is the moduli space of flat connections on $M$ \cite{WittenOldpaper}.)

This discussion has been somewhat abstract. For a concrete example, we can consider the purely fermionic manifold $\mathbb{C}\mathbb{P}^{0|2}$ which has the symplectic form
\begin{align}
\omega &= \partial \bar{\partial}\log(1 + \theta^1\bar{\theta}^1 + \theta^2\bar{\theta}^2)= -(1 - \theta^2\bar{\theta}^2)d\theta^1d\bar{\theta}^1 - \theta^1\bar{\theta}^2d\theta^2d\bar{\theta}^1 + (1\leftrightarrow 2).
\end{align}
We would like to consider the integral
\be
I = \int d^2\theta^1 d^2\theta^2d^2(d\theta^1)d^2(d\theta^2) \exp\left(\frac{1}{2}\omega_{ij}d\theta^id\theta^j + \frac{1}{g^2} H\right)
\ee
where $H$ is a symplectic generator of a $U(1)$. For example
\be
H = \theta^1\bar{\theta}^1 - \theta^1\bar{\theta}^1\theta^2\bar{\theta}^2 + (1\leftrightarrow 2)
\ee
which phase rotates $\theta^1,\bar{\theta}^1$ into each other and similarly $\theta^2,\bar{\theta}^2$. The Gaussian integral over the bosonic $d\theta$ variables gives us a measure factor proportional to $(1 + \theta^1\bar{\theta}^1 + \theta^2\bar{\theta}^2)$. We then do the final integral over the $\theta$ variables. This gives that $I$ is a multiple of $1/g^4$. This is the same as the ``one-loop'' answer where we replace everything by the lowest nonvanishing order in the $\theta$ variables. The fact that this works relies on a cancellation between a measure factor and a factor coming from the higher order term in $H$. This is similar to what happens in bosonic examples of DH integrals.

\section{A Lattice Version of the Schwarzian Theory}\label{lattice}
In this appendix, we discuss a lattice regularization of the Schwarzian path integral that preserves the exact $SL(2,\R )$ gauge symmetry of the Schwarzian action.\footnote{We are grateful to J.~Maldacena for suggesting this lattice implementation.  Note that we do
not expect the regularized theory to be one-loop exact.} This was not necessary for anything in the main text, but we include it in order to give a concrete picture of the important $\phi(\tau)$ configurations at different values of the coupling. We expect that the lattice theory is only one-loop exact in the continuum limit.

One can work in terms of the variable $f(\tau) = \tan\frac{\phi(\tau)}{2}$ or in terms of $\phi(\tau)$ directly. We start first with $f$. One approximates the continuum $f(\tau)$ by a lattice of values $f_i$ at $\tau=ai$, $i=0,1,2\dots$. Here we take the lattice spacing to be $a$ and the inverse temperature to be $2\pi$, so we have $n = \frac{2\pi}{a}$ points. Then one can make a lattice approximation to the Schwarzian using a conformal cross ratio of adjacent values:
\be
\frac{(f_{i+3}-f_{i+1})(f_{i+2}-f_i)}{(f_{i+3}-f_{i+2})(f_{i+1}-f_i)} = 4 - 2a^2 \text{Sch}(f,u) + O(a^3).
\ee
This preserves an exact $SL(2,\R )$ acting on all points as $f \rightarrow \frac{a f+b}{cf+d}$. So a lattice version of the Schwarzian theory, including a lattice version of the measure $\frac{df}{f'}$ from (\ref{naiveM}), is
\be
Z(g) = \int \left(\prod_{i = 1}^{n} \frac{df_i}{f_{i+1} - f_i}\right)e^{-\frac{1}{2ag^2} \sum_{i=1}^n \left[\frac{(f_{i+3}-f_{i_1})(f_{i+2}-f_i)}{(f_{i+3}-f_{i+2})(f_{i+1}-f_i)}-4\right]}.
\ee
In these formulas the index should be understood to ``wrap around'' the circle, so that $i = n+1$ is equivalent to $i = 1$. Notice that our measure is also exactly $SL(2,\R )$ invariant.

\begin{figure}[t]
\begin{center}
\includegraphics[scale = 0.43]{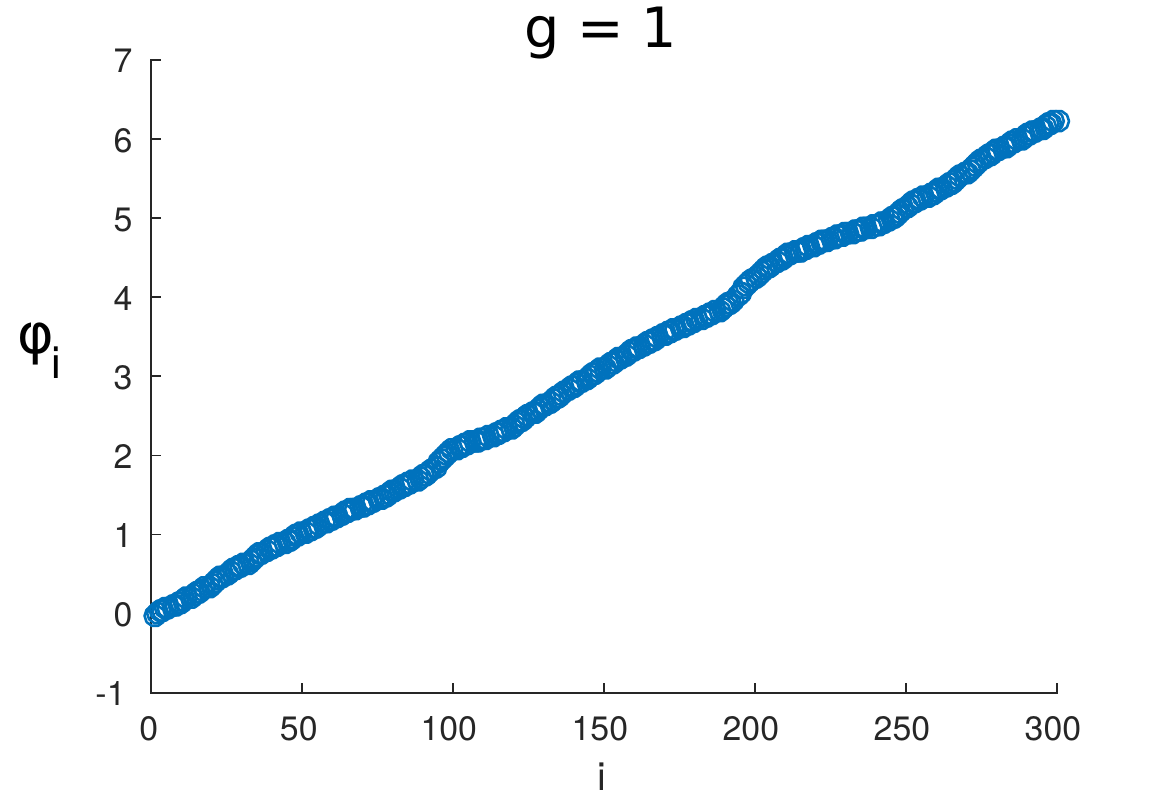}
\includegraphics[scale = 0.43]{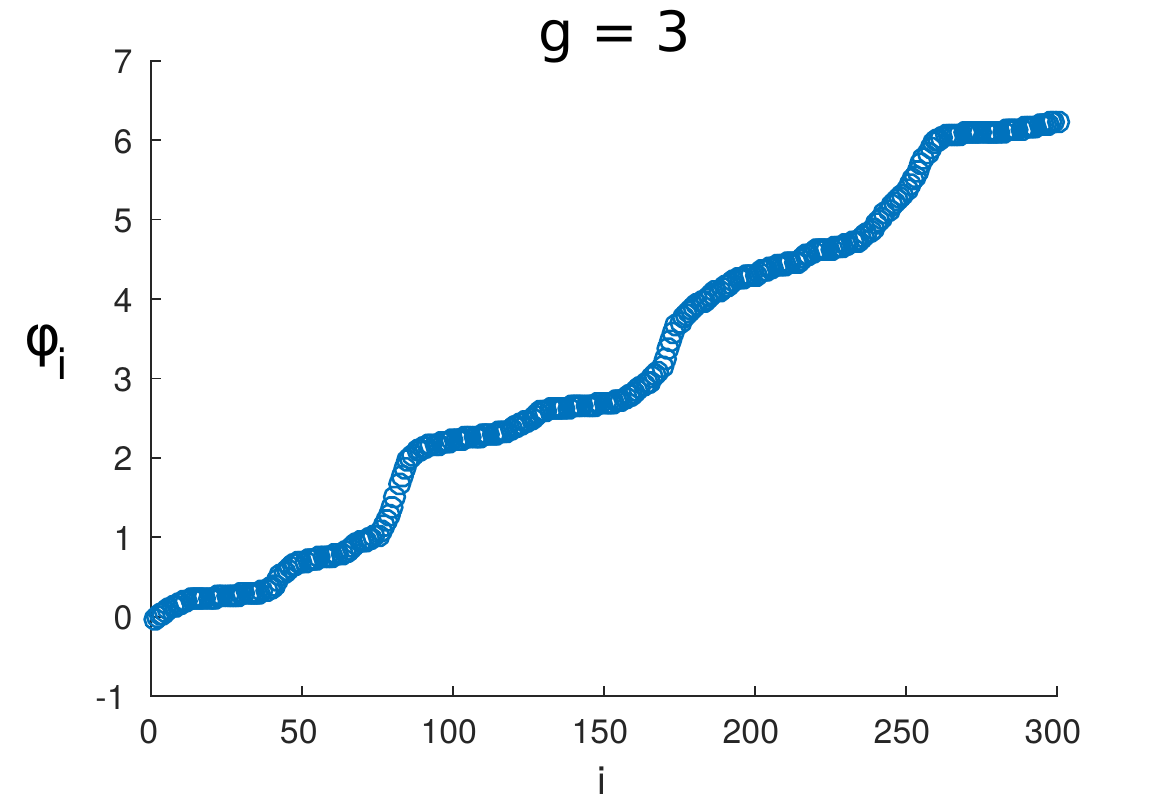}
\includegraphics[scale = 0.43]{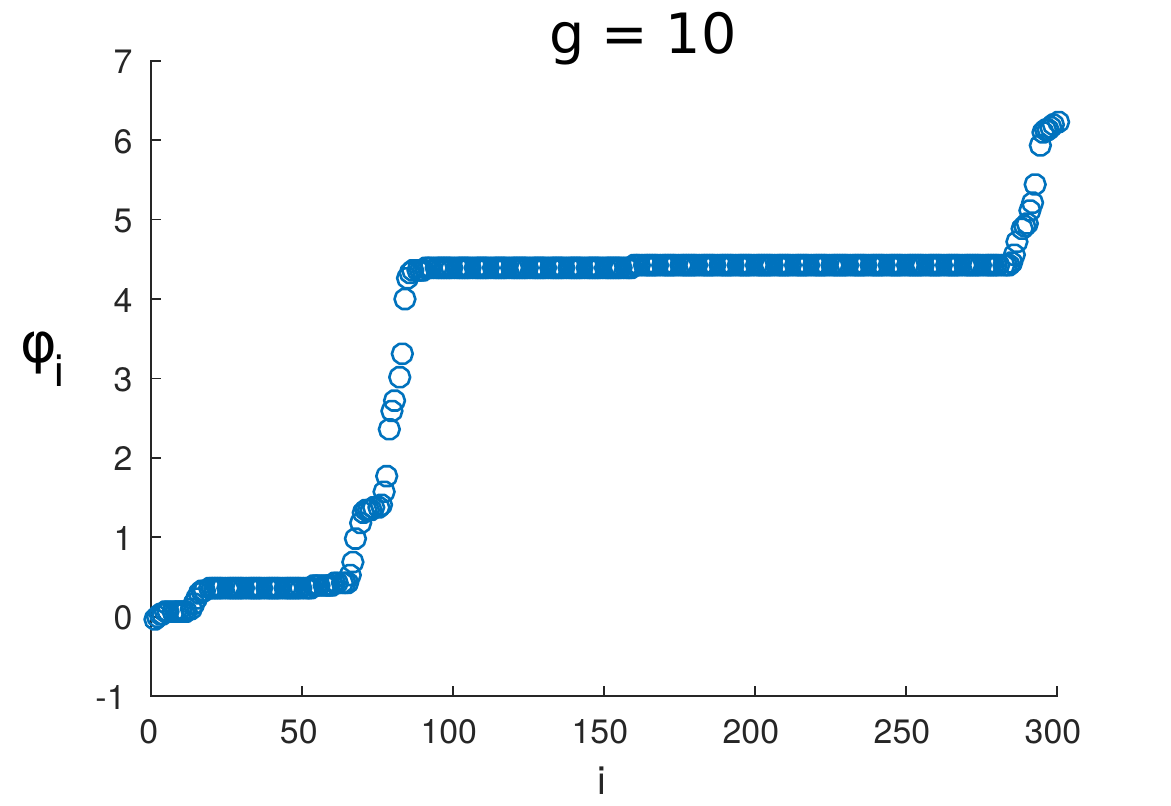}
\caption{Typical configurations of the discretized $\phi$ varible that contribute to $Z(g)$ for the discretized Schwarzian theory. The configurations were generated using the Metropolis algorithm, with a lattice of $n = 300$. The figure at left contains only small fluctuations about the saddle point, which would be a straight line $\phi_j = \frac{2\pi j}{n}$.}\label{MC}
\end{center}
\end{figure}
We can also directly use the $\phi$ variable instead of $f$. Then the partition function is
\be
Z(g) = \int \left(\prod_{i = 1}^{n} \frac{d\phi_i}{\sin\frac{\phi_{i+1} - \phi_i}{2}}\right)e^{-\frac{1}{2a g^2} \sum_{i=1}^n \left[\frac{\sin\frac{\phi_{i+3}-\phi_{i_1}}{2}\sin\frac{\phi_{i+2}-\phi_i}{2}}{\sin\frac{\phi_{i+3}-\phi_{i+2}}{2}\sin\frac{\phi_{i+1}-\phi_i}{2}}-4\right]}.
\ee
This reduces to the expected thing once we realize
\be
\frac{\sin\frac{\phi_{i+3}-\phi_{i_1}}{2}\sin\frac{\phi_{i+2}-\phi_i}{2}}{\sin\frac{\phi_{i+3}-\phi_{i+2}}{2}\sin\frac{\phi_{i+1}-\phi_i}{2}}= 4 - 2a^2\left[\text{Sch}(\phi,u) + \frac{1}{2}\phi'^2\right] + O(a^3).
\ee
Both of the above partition functions are divergent because of the exact $SL(2,\R )$ gauge symmetry. We can fix this by fixing the values of three of the points. One choice is to fix $\phi_1 = 0, \phi_2 = \frac{2\pi}{n}, \phi_N = 2\pi(1 - \frac{1}{n})$. Then the integral over monotonic $\phi$ is convergent. Notice that the log divergence from the measure is cut off by the fact that the action diverges when neighboring values of $\phi$ coincide. 

Figure \ref{MC} shows some typical configurations contributing to $Z(g)$, sampled using the Metropolis algorithm, for different values of $g$ and with $n = 300$ points on the lattice. As $g$ increases, the free energy becomes increasingly dominated by the measure, which prefers to have all of the points at very similar values. Of course, the function $\phi_i$ is required to run between zero and $2\pi$, but for large $g$ it prefers to do this all at once in a few places. 

\section{Correlation Functions of the Schwarzian}
In this appendix, we show how to compute correlation functions of the Schwarzian derivative of $\phi$. To start, one can compute moments of the integrated Schwarzian by differentiating the partition function. For example:
\begin{align}\label{zpp1}
\int d\tau \langle \sch(\tau)\rangle &= \frac{1}{Z(g)}\frac{\partial}{\partial_{1/g^2}}{Z(g)} = \pi + \frac{3g^2}{2}\\
\int d\tau d\tau' \langle \sch(\tau)\sch(\tau')\rangle &= \frac{1}{Z(g)}\frac{\partial^2}{\partial_{1/g^2}^2}{Z(g)}= \pi^2 + 3\pi g^2 + \frac{3 g^4}{4}.\label{zpp}
\end{align}
Here and below, we are using the notation $\sch(\tau) \equiv \sch(\tan\frac{\phi(\tau)}{2},\tau)$. To get unintegrated 
correlators at fixed points, we proceed as follows. The composition rule
\be
\sch(f(y(x)),x) = \sch(f,y)(\partial_x y)^2 + \sch(y,x)
\ee
together with invariance of the measure implies
\be
Z(g) = \int \frac{d\mu[\phi]}{SL(2)}e^{\frac{1}{g^2}\int dx \sch(\tan\frac{\phi(y)}{2},x)} = \int \frac{d\mu[\phi]}{SL(2)}e^{\frac{1}{g^2}\int dx \left[(\partial_x y)^2\sch(\tan\frac{\phi}{2},y) + \sch(y,x)\right]}.
\ee
Relabeling $x\rightarrow h$ and $y\rightarrow \tau$, and using $\sch(y,x) = -(\partial_x y)^2\sch(x,y)$, we find
\be\label{bnm}
\int \frac{d\mu[\phi]}{SL(2)}e^{\frac{1}{g^2}\int \frac{d\tau}{h'(\tau)}\sch(\tan\frac{\phi(\tau)}{2},\tau)} = e^{\frac{1}{g^2}\int \frac{d\tau}{h'(\tau)}\, \sch(h,\tau)}Z(g) = e^{\frac{1}{2g^2}\int d\tau\frac{h''(\tau)^2}{h'(\tau)^3}}Z(g).
\ee
Here, we require that $h$ is a function that maps $S^1\rightarrow S^1$,
\be
h(\tau) = \tau + \epsilon(\tau),\hspace{20pt} \int \epsilon'(\tau)d\tau = 0.
\ee
Expanding (\ref{bnm}) to second order in $\epsilon$ we find
\be
\langle \sch(\tau)\sch(0)\rangle = -2g^2\langle \sch(0)\rangle\delta(\tau) - g^2\delta''(\tau) + \text{const.}
\ee
The value $\langle \sch(0)\rangle$ can be determined from (\ref{zpp1}) and the constant can be determined using the integrated correlator (\ref{zpp}). The result is 
\begin{align}
\langle \sch(\tau)\sch(0)\rangle &= -\left(g^2 + \frac{3 g^4}{2\pi}\right)\delta(\tau) - g^2\delta''(\tau) + \frac{1}{4}  + \frac{5g^2}{4\pi} + \frac{15g^4}{16\pi^2}.
\end{align}
Higher point correlation functions of the Schwarzian can be evaluated in a similar way. We will point out two interesting features. First, the correlators are constant away from coincident points. This can be understood from the fact that the equation of motion for $\phi$ is $\sch'(\tau) = 0$. Second, the actual values of the separated-point correlators are given by the corresponding moments of the energy of the system and the relation
\begin{align}
E = \frac{2\pi}{\beta g^2}\sch(\tau).
\end{align}
More precisely, what we mean is that the separate-points correlators of the Schwarzian give moments of the energy in the distribution $\rho(E)$. These can be computed in the usual way by differentiating (\ref{partbeta}). We have checked this mechanically up to the three-point function, but we expect that it holds in general.

\mciteSetMidEndSepPunct{}{\ifmciteBstWouldAddEndPunct.\else\fi}{\relax}
\bibliographystyle{utphys}
\bibliography{refs.bib}{}

\end{document}